\documentclass[11pt]{article}
\usepackage{amssymb}
\usepackage{graphics}
\usepackage{epsfig}
\usepackage{a4wide}

\begin{document}

\title{Full one-loop electroweak corrections to $h^0(H^0,A^0) H^\pm W^\mp$
associated productions at $e^+e^-$ linear
colliders\footnote{Supported by National Natural Science
Foundation of China.}} \vspace{3mm}

\author{{ Liu Jing-Jing, Ma Wen-Gan, Zhang Ren-You, Guo Lei, Jiang Yi, Han Liang}\\
{\small  Department of Modern Physics, University of Science and Technology}\\
{\small  of China (USTC), Hefei, Anhui 230027, People's Republic
of China}\\  }

\date{}
\maketitle \vskip 12mm

\begin{abstract}
\par
We study the complete one-loop electroweak(EW) corrections to the
processes of single charged Higgs boson production associated with
a neutral Higgs boson$(h^0,H^0,A^0)$ and a gauge boson $W^\pm$ in
the framework of the minimal supersymmetric standard model(MSSM).
Numerical results at the ${\rm SPS1a'}$ benchmark point as
proposed in the SPA project, are presented for demonstration. We
find that for the process $e^+e^-\to h^0H^\pm W^\mp$ the EW
relative correction can be either positive or negative and in the
range of $-15\%\sim 20\%$ in our chosen parameter space. While for
the processes $e^+e^-\to H^0(A^0)H^\pm W^\mp$ the corrections
generally reduce the Born cross sections and the EW relative
corrections are typically of order $-10\%\sim -20\%$.
\end{abstract}

\vskip 5cm {\large\bf PACS: 12.15.LK, 12.60.Jv, 14.70.Fm,
14.80.Cp}


\baselineskip=0.3 in

\newpage
\newcommand{\nb}{\nonumber}
\section{Intorduction}
\par
Even though the Standard Model(SM)\cite{sm1,sm2} has been
successful in describing the strong, weak and electromagnetic
interaction phenomena at the energy scale up to $10^2$ GeV,  it
has been known for a long time that it is still considered as a
low-energy effective theory suffering from a number of theoretical
difficulties, such as hierarchy problem. It is likely that at the
higher energy scale the minimal supersymmetric standard model
(MSSM)\cite{mssm1,mssm2,mssm3} is the most attractive candidate
among various extensions of the SM, which is expected to resolve
theoretical difficulties existing in the SM. By adopting two Higgs
doublets to preserve the supersymmetry in the MSSM, five Higgs
bosons are predicted. They are two CP-even Higgs scalar bosons
$h^0$ and $H^0$, one CP-odd Higgs boson $A^0$ and a charged Higgs
pair $H^\pm$.

\par
The charged Higgs boson plays an important role in the extensions
of the SM, its discovery in experiment will unequivocally imply
the existence of the physics beyond the SM. If the charged Higgs
boson really exists in nature then it is mostly possible to be
discovered in the coming few years at the CERN Large Hadron
collider (LHC) due to its TeV scale colliding energy, but the
further accurate measurements of masses of Higgs bosons and
investigations of the properties of Higgs bosons\cite{LC}, which
is crucial for verification of electroweak (EW) symmetry breaking
mechanism, need be best performed in a clean environment of
$e^+e^-$ collisions, such as at the future International Linear
Collider (ILC) \cite{ILC}. The study would be of the utmost
importance for establishment of the MSSM or some of more general
supersymmetric models as the theory of the EW interactions. In
addition, further study of the complete Higgs spectrum is the
cornerstone of the definitive test of mechanism of electroweak
symmetry breaking. In discovery and detailed study of the
properties for these Higgs bosons, there are some important
processes for neutral and charged Higgs bosons at linear
colliders, such as, $e^+e^-\to Z^*\to h^0(H^0) Z^0$, $e^+e^-\to
Z^*\to h^0(H^0) A^0$, $e^+e^-\to \nu \bar{\nu}W^{+*} W^{-*}\to \nu
\bar{\nu} h^0(H^0)$, $e^+e^-\to H^+H^-$ and so on\cite{komamiya}.
The lightest neutral Higgs boson $h^0$ in the MSSM has the mass
upper limit of about $140 ~\rm{GeV}$, and it decays dominantly
into $b\bar{b}$ pairs. For the heavier neutral Higgs bosons
$H^0/A^0$, as a result of the strong enhancement of the coupling
between Higgs boson and down-type fermions, both $H^0$ and $A^0$
will decay almost exclusively into $b\bar{b}$ ($\sim 90\%$) and
$\tau^+\tau^-$ ($\sim 10\%$) pairs (The decays $H^0, A^0 \to
t\bar{t}$ are possible only when they are kinematically allowed),
and on the contrary all other decays are strongly
suppressed\cite{Carena,Janot,Sopczak,Stirling,Djouadi}. For the
mass of charged Higgs boson being blow $170~{\rm GeV}-180{\rm
GeV}$ (i.e., the $H^- \to \bar tb$ decay threshold), the dominant
decay mode of the charged Higgs boson is $H^- \to
\tau^-\bar{\nu}_\tau$. But when the mass of charged Higgs boson is
above the $H^- \to \bar tb$ decay threshold, the charged $H^\pm$
boson mainly decays into $tb$ pair leading to $Wb\bar{b}$ final
states\cite{Moretti}.

\par
If $H^+H^-$ pair production at $e^+e^-$ linear colliders(LC) is
kinematically allowed, the dominant production process of charged
Higgs bosons is the pair production via exchanging a virtual
photon and $Z$ boson $e^+e^-\to \gamma^*,Z^*\to
H^+H^-$\cite{komamiya}. For a fixed $e^+e^-$ colliding energy, its
cross section is related only with the charged Higgs boson mass.
Another important production mode is the production of single
charged Higgs boson, especially when the charged Higgs boson is
too heavy to be produced in pair at linear colliders. Among
various single charged Higgs boson production channels, the bottom
quark associated production process of $e^+e^-\to b\bar{b}H^\pm
W^\mp$\cite{Moretti} is experimentally interesting since it can be
yielded by many topologies, such as the $t\bar t$ and $H^+H^-$
pair productions followed by their corresponding sequential
decays. The $e^+e^-\to b\bar{b}H^\pm W^\mp$ process can be also
induced from following three single charged Higgs boson production
processes (\ref{process}) and the sequential decays $h^0(H^0,A^0)
\to b\bar b$,
\begin{eqnarray} \label{process}
 e^+e^-\to \Phi^0 H^\pm W^\mp, ~~~(\Phi^0=h^0,
H^0,A^0).
\end{eqnarray}

S. Kanemura, S. Moretti, {\it et al,} studied the processes of
single charged Higgs boson production channels in the MSSM at LCs,
including the processes (\ref{process}) at tree-level in
Ref.\cite{Moretti,Kanemura}. It was found that there are regions
of parameter space where it is accessible beyond the kinematic
limit for pair production, and these processes are excellent
channels in studying the interactions among neutral Higgs
boson($h^0$, $H^0$ or $A^0$), charged Higgs boson and gauge
boson($W^\pm$), and is a useful alternative channel for studying
the phenomenology of charged Higgs boson.

\par
In present work, we will study the EW corrections to the processes
(\ref{process}) including complete one-loop diagrams in the MSSM.
The paper is organized as follows: In Sect.2, we present the
analytical calculation of the complete one-loop EW radiative
corrections to $e^+e^- \to \Phi^0 H^\pm W^\mp(\Phi^0=h^0,H^0,A^0)$
processes in the MSSM as well as the treatment of charged Higgs
boson resonance. The numerical results and discussions are given
in Sect.3. Finally, we give a short summary.

\par
\section{Calculations}
\subsection{Conventions and notations in analytical calculations}
\par
We denote the processes as
\begin{eqnarray}
\label{process1} e^+(p_1)+e^-(p_2)\to \Phi^0(p_3)+ H^+(p_4)
+W^-(p_5),~~~(\Phi^0=h^0, H^0, A^0).
\end{eqnarray}
The lowest order cross section reads
\begin{eqnarray}
\sigma_{tree}= \frac{(2 \pi )^4}{4|\vec{p}_1|\sqrt{s}}\int
\overline{\sum_{spin}} |{\cal M}_{tree}|^2 d\Phi_3
\end{eqnarray}
where $\vec{p}_1$ is the c.m.s. kinematical momentum of the
incoming positron, The integration is performed over the
three-body phase space of final particles $\Phi^0 H^+W^{-}$, the
bar over summation indicates averaging over initial spins. The
phase-space element $d\Phi_3$ is defined by
\begin{eqnarray}
{d\Phi_3}=\delta^{(4)} \left( p_1+p_2-\sum_{i=3}^5 p_i \right)
\prod_{j=3}^5 \frac{d^3 \textbf{\textsl{p}}_j}{(2 \pi)^3 2 E_j}
\end{eqnarray}

\par
In our calculation, we use dimensional reduction(DR)
regularization scheme, which is supersymmetric invariant at least
one-loop level, to regularize UV divergences and adopt
on-shell(OS) conditions(neglecting the finite widths of the
particle) to renormalize the relevant fields\cite{ren}. We adopt
the 't Hooft-Feynman gauge and the definitions of one-loop
integral functions in Ref.\cite{loop} in the calculation of
one-loop diagrams. The Feynman diagrams and their corresponding
amplitudes are created by $FeynArts~3$ \cite{Feynarts}
automatically, and the Feynman amplitudes are subsequently reduced
by $FORM$\cite{Form}. We depict the lowest-order Feynman diagrams
in Fig.1. The numerical calculation of the two-, three- and
four-point integral functions are done by using FF
package\cite{van}. The implementations of the scalar and the
tensor five-point integrals are carried out exactly by using the
Fortran programs as used in our previous works\cite{youyu} with
the approach presented in Ref.\cite{pentagon}. Due to the fact
that the Yukawa coupling of Higgs/Goldstone to the fermions is
proportional to the mass of the fermion, we ignore the
contributions of the Feynman diagrams which involve the Yukawa
coupling between any Higgs/Goldstone boson and electrons.

\begin{figure}[htp]
\centering
\includegraphics[scale=0.8,bb=39 367 558 600]{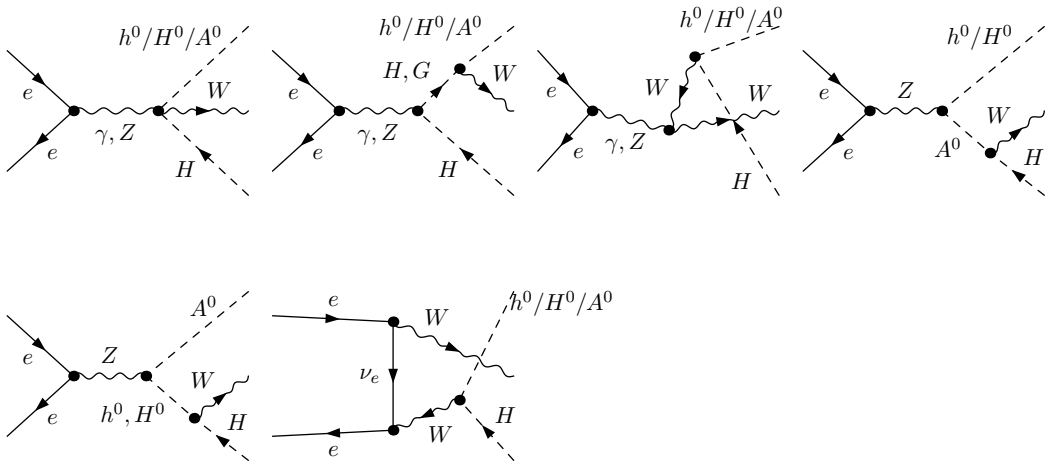}
\caption{The tree-level Feynman diagrams of the processes $e^+e^-
\to \Phi^0 H^+ W^-(\Phi^0=h^0,H^0,A^0)$.}
\end{figure}

\par
The ${\cal O}(\alpha_{ew})$ virtual correction to the cross
section can be expressed as
\begin{eqnarray}
\sigma_{virtual}=\sigma_{tree}\delta_{virtual}= \frac{(2 \pi
)^4}{2|\vec{p}_1|\sqrt{s}}\int d\Phi_3\overline{\sum_{spin}}
Re({\cal M}_{tree}{\cal M}^{\dag}_{virtual})
\end{eqnarray}
where $\mathcal{M}_{\mathrm{virtual}}$ is the renormalized
amplitude involving all the one-loop electroweak Feynman diagrams
and corresponding counter-terms\cite{RevD}.

\par
The definitions and the explicit expressions of these
renormalization constants can be found in
Refs.\cite{MSSM1,MSSM2,eetth}. Here we just list the definitions,
which do not appear in Ref.\cite{eetth}, as below:
\begin{eqnarray}
& & \left( \begin{array}{c} G^{+(0)} \\ H^{+(0)}
\end{array}
\right)=\left(
\begin{array}{cc}
1 + \frac{1}{2} \delta Z_{G^+G^+} & \frac{1}{2} \delta Z_{G^+H^+} \\
\frac{1}{2} \delta Z_{H^+G^+}  & 1 + \frac{1}{2} \delta Z_{H^+H^+}
\end{array}
\right) \left(
\begin{array}{c}
G^+ \\ H^+
\end{array}
\right),\nonumber\\
& &M_{h^0}^{({0})2} =M_{h^0}^2 + \delta M_{h^0}^2,~~~~~~~~~~~~~~
M_{H^0}^{({0})2}~=~ M_{H^0}^2 + \delta M_{H^0}^2,\nb \\
& &M_{H^+}^{({0})2} = M_{H^+}^2 + \delta M_{H^+}^2,~~~~~~~~~~~~
\delta M_{h^0}^2 =
\widetilde{Re}\left[\Sigma^{h^0h^0}(M_{h^0}^2)-b_{h^0h^0}\right],\nb\\
& &\delta M_{H^0}^2 =
\widetilde{Re}\left[\Sigma^{H^0H^0}(M_{H^0}^2)-b_{H^0H^0}\right],~~~~~~~
\delta M_{H^+}^2 =
\widetilde{Re}\left[\Sigma^{H^+H^+}(M_{H^+}^2)-b_{A^0A^0}\right], \nb\\
& &\delta Z_{H^+H^+}=-\widetilde{Re} \frac{\partial \Sigma^{H^+
H^+}(k^2)}{\partial k^2} \mid_{k^2 = M_{H^+}^2},~~~~~ \delta
Z_{G^+H^+}=\frac{2}{M_{H^+}^2} \widetilde{Re} [ b_{G^0A^0} -
\Sigma^{G^+H^+}(m_{H^+}^2)],\nb \\
& &\delta Z_{H^+G^+}=-\frac{2}{M_{H^+}^2} \widetilde{Re} [
b_{G^0A^0} - \Sigma^{H^+G^+}(0)], ~~\delta Z_{G^+G^+}=
-\widetilde{Re} \frac{\partial \Sigma^{G^+
G^+}(k^2)}{\partial k^2} \mid_{k^2 =0}.\nb \\
\end{eqnarray}
The Higgs tadpole parameters $ b_{A^0A^0}$, $b_{G^0A^0}$,
$b_{h^0h^0}$, $b_{H^0h^0}$ and $b_{H^0H^0}$ are defined and
expressed as in Ref.\cite{MSSM2}. The operator $\widetilde{Re}$
takes only the real part of the loop integrals and does not affect
the possible complex couplings.

\subsection{Disposal of the charged Higgs boson resonance}
\begin{figure}[htp]
\centering
\includegraphics[scale=0.8]{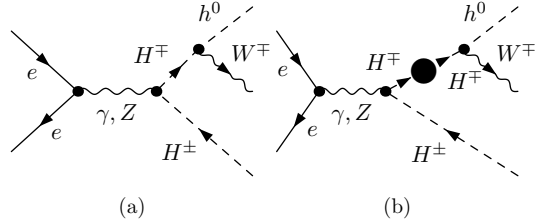}
\caption{Feynman diagrams with charged Higgs boson resonance for
process $e^+e^-\to h^0 H^+ W^-$}
\end{figure}

\par
The disposal of $H^-$ resonance in the calculation of the EW
correction is similar with the method described in
Ref.\cite{width}. For the process $e^+e^- \to h^0 H^+ W^-$ in
$TeV$ scale colliding energy, the amplitude of the Feynman diagram
with internal line of $H^-$(shown in Fig.2(a), denoted as ${\cal
M}_{tree}^{H^-}(\Gamma_{H^-}=0)$) suffers from resonance effect
when $s_{34}\equiv(p_3+p_4)^2 \to M_{H^-}^2$ due to the real
charged Higgs boson decay $H^- \to h^0 W^-$. We have to introduce
the complex $H^-$ mass to do the following replacement in the
resonance propagator of the amplitude ${\cal
M}_{tree}^{H^-}(\Gamma_{H^-}=0)$,
\begin{eqnarray}
\frac{1}{s_{34}-M_{H^-}^2} \to \frac{1}{s_{34}-M_{H^-}^2+i {\cal
M}_{H^-}\Gamma_{H^-}}, \label{Replace}
\end{eqnarray}
where $\Gamma_{H^-}$ is $H^-$ decay width. Then we get the
amplitude at the tree-level as
\begin{eqnarray}
{\cal M}_{tree}^{H^-}=\frac{s_{34}-M_{H^-}^2}{s_{34}-M_{H^-}^2+i
{\cal M}_{H^-}\Gamma_{H^-}}{\cal M}_{tree}^{H^-}(\Gamma_{H^-}=0).
\end{eqnarray}

The modified amplitude ${\cal M}_{tree}^{H^-}$ is safe amplitude
being free of the $H^-$ resonance singularity. We know that at the
tree-level calculation this replacement does not induce the gauge
invariant and double-counting problems\cite{width}, but we should
remember that after this treatment, there includes the
contribution from the imaginary part of charged Higgs boson
self-energy($\Sigma^{H^-H^-}$) at higher order. In order to avoid
the double-counting in calculation up to one-loop level, we should
subtract the higher order contribution from the tree-level
amplitude ${\cal M}_{tree}^{H^-}$. In our practical calculation,
we separate some special Feynman diagrams(shown in Fig.2(b)) from
entirety. Each of these diagrams in Fig.2(b) contains a one-loop
of charged Higgs boson self-energy, or the corresponding
counterterm. Then we can express the one-loop level amplitude as
below:
\begin{eqnarray}
{\cal M}^{fix-width}_{one-loop}&=&\left[{\cal
M}_{one-loop}(\Gamma_{H^-}=0)-{\cal
M}_{H^-H^--self}(\Gamma_{H^-}=0)
\right]_{\frac{1}{S_{34}-M_{H^-}^2} \to
\frac{1}{S_{34}-M_{H^-}^2+i M_{H^-}\Gamma_{H^-}}} \nb\\
& &
-\left[\frac{\Sigma^{H^-H^-}(s_{34})-\left[\Sigma^{H^-H^-}(M_{H^-}^2)-b_{A^0A^0}\right]}{s_{34}-M_{H^-}^2}
+\delta Z_{H^-H^-} \right]{\cal M}_{tree}^{H^-},
\end{eqnarray}
where the first term on right-handed side is just the amplitude at
one-loop level after subtracting the the contributions of
Fig.2(b). The one-loop(or higher) order contributions are included
by doing the replacement shown in Eq.(\ref{Replace}). The real
part of ${\Sigma^{H^-H^-}(M_{H^-}^2)-b_{A^0A^0}}$ is just equal to
$\delta M_{H^-}^2$, the mass counterterm of charged Higgs boson,
while ${\Sigma^{H^-H^-}(M_{H^-}^2)-b_{A^0A^0}}$ receives its
imaginary part in the resonance from the subtraction of the $i
M_{H^-}\Gamma_{H^-}$ contribution already contained in ${\cal
M}_{tree}^{H^-}$. In the calculation, charged Higgs boson width
was evaluated using the HDECAY package\cite{HDECAY}, and UV
divergence cancelled numerically.

\subsection{Real photonic corrections}
\par
The ${\cal O}(\alpha_{ew})$ virtual correction suffers from both
ultraviolet(UV) and infrared(IR) divergences. It is checked
numerically that the UV divergence vanished after renormalization.
IR divergence comes from virtual photonic loop correction. It can
be cancelled by the contribution of the real photon emission
processes. We denote the real photon emission as
\begin{eqnarray}
\label{soft} e^+(p_1)+e^-(p_2)\to \Phi^0(p_3)+ H^+(p_4) +W^-(p_5)+
\gamma(k_\gamma),~~(\Phi^0=h^0,H^0,A^0).
\end{eqnarray}
We adopt the general phase-space-slicing method \cite{PSS} to
separate the soft photon emission singularity from the real photon
emission processes. By using this method, the bremsstrahlung phase
space is divided into singular and non-singular regions by the
soft photon cutoff $\Delta E_{\gamma}$, then the relative
correction of the real photon emission is broken down into
corresponding soft term $\delta_{soft}$ with radiated photon
energy $k^0_\gamma<\Delta E_{\gamma}$ and hard terms
$\delta_{hard}$ with $k^0_\gamma>\Delta E_{\gamma}$, where
$k^0_\gamma=\sqrt{ |\vec{k}_{\gamma}|^2+ m_{\gamma}^2} $ and
$m_\gamma$ is the mass of photon, which is used to regulate the IR
divergences existing in the soft term. The IR divergence from the
soft contribution cancels that from the virtual correction
exactly\cite{Hooft,Denner}. Therefore, the sum of the virtual and
soft cross section is independent of the IR regulator $m_\gamma$.
In the calculation, The phase space integration of the processes
$e^+e^-\to \Phi^0 H^+ W^- \gamma,(\Phi^0=h^0,H^0,A^0)$ with hard
photon emission is performed by using the program $GRACE$
\cite{grace}. The hard photon emission cross section
$\sigma_{hard}$ is UV and IR finite with $k_\gamma>\Delta
E_\gamma$. Finally, we can get the UV and IR finite ${\cal
O}(\alpha_{ew})$ correction.

\par
\section{Numerical Results}
\par
In the numerical calculation, we use the following SM parameters
\cite{RevD}
\begin{eqnarray}
m_e&=&0.510998902~{\rm MeV},~m_\mu~=~105.658369~{\rm MeV},~~m_\tau~=~1776.99~{\rm MeV},\nb\\
m_u&=&66~{\rm MeV},~~~~~~~~~~~~~~m_c~=~1.2~{\rm GeV},~~~~~~~~~~~~~m_t~=~178~{\rm GeV},\nb\\
m_d&=&66~{\rm MeV},~~~~~~~~~~~~~~m_s~=~150~{\rm MeV},~~~~~~~~~~~~m_b~=~4.7~{\rm GeV} ,\nb\\
m_W&=&80.425~{\rm GeV},~~~~~~~~~m_Z~=~91.1876~{\rm GeV},
~~~~~~~\alpha_{ew}(0)~=~1/137.036.
\end{eqnarray}
Here we take the effective values of the light quark masses $(m_u$
and $m_d)$ which can reproduce the hadronic contribution to the
shift in the fine structure constant $\alpha_{ew}(m_Z^2)$
\cite{DESY}. As a numerical demonstration, we generally adopt the
relevant supersymmetric parameters from the CP and R-parity
invariant MSSM reference point ${\rm SPS1a'}$ which is defined in
the SPA project\cite{SPA,snowmass}. This point satisfies all the
available high- and low-energy precision data and both the bounds
for the masses of the SUSY particles and the bounds from
cosmology. As we use the on-shell renormalization scheme, we have
to transform the ${\rm SPS1a'}$ $\overline{DR}$ input parameters
${\cal P}$ into on-shell parameters ${\cal P}^{\rm OS}$. This
transformation is simply performed by subtracting the
corresponding counterterms i.~e. $ {\cal P}^{\rm OS} = {\cal
P}(Q)-\delta {\cal P}(Q)$. The input parameters for the reference
point ${\rm SPS1a'}$ are defined in the $\overline{DR}$ scheme at
the scale $Q=1~TeV$, which are a set of mSUGRA parameters listed
below:
\begin{eqnarray}
M_{1/2}=250~GeV,~~M_0=70~GeV,~~A_0=-300~GeV, \\
sign(\mu)=+1,~~\tan\beta(\tilde{M}=1~TeV)=10.
\end{eqnarray}
The relevant parameters of ${\rm SPS1a'}$ point are listed in
Table~\ref{table}.

\begin{table}
\center
\begin{tabular}{ |c|c||c|c|  }
\hline
Particle & Mass(GeV) & Particle & Mass(GeV) \\
\hline \hline
$h^0$              & 116.0 & $\tilde{u}_R$        & 547.2\\
$H^0$              & 425.0 & $\tilde{u}_L$        & 564.7\\
$A^0$              & 424.9 & $\tilde{d}_R$        & 546.9\\
$H^+$              & 432.7 & $\tilde{d}_R$        & 570.1\\
 \hline
$\tilde{e}_R$      & 125.3 & $\tilde{\tau}_1$     &  107.9\\
$\tilde{e}_L$      & 189.9 & $\tilde{\tau}_2$     & 194.9\\
$\tilde{\nu}_e$    & 172.5 & $\tilde{\nu}_{\tau}$ & 170.5 \\
\hline
$\tilde{\chi}_1^0$ & 97.7  & $\tilde{t}_1$        & 366.5\\
$\tilde{\chi}_2^0$ & 183.9 & $\tilde{t}_2$        & 585.5\\
$\tilde{\chi}_3^0$ & 400.5 & $\tilde{b}_1$        & 506.3\\
$\tilde{\chi}_4^0$ & 413.9 & $\tilde{b}_2$        & 545.7\\
\hline
$\tilde{\chi}_1^+$ & 183.7 &  &  \\
$\tilde{\chi}_2^+$ & 415.4 &  &  \\
\hline
\end{tabular}
\caption{Mass spectrum of supersymmetric particles and Higgs
bosons in the reference point $SPS1a'$. }\label{table}
\end{table}

\par
Since the difference between $\overline{DR}$ and OS mass values
for sparticle and Higgs bosons is of high order, either of them
can be used. In our paper we generally adopt the $\overline{DR}$
masses of sparticle and Higgs boson, but when we study the cross
sections as the functions of $M_{A^0}$ or take $M_{A^0}\neq 424.9
~GeV$, we still adopt the ${\rm SPS1a'}$ MSSM spectrum and
parameters except Higgs sector. In the later case the masses of
the MSSM Higgs bosons are fixed by taking into account the
significant radiative corrections in the MSSM by using FormCalc
package with the input parameters: $M_{A^0}$ and OS $\tan \beta$
value(i.e., $\tan\beta=10.31$ at ${\rm SPS1a'}$
point)\cite{FeyArtsFormCalc}. We checked in our calculation that
the total cross section is independent on the IR regulator
parameter $m_\gamma$ and the soft cutoff $\delta_s$. In following
we take $m_{\gamma}=10^{-2}~GeV$ and $\delta_s=10^{-4}$.

\par
The Born and ${\cal O}(\alpha_{ew})$ electroweak corrected cross
sections at the ${\rm SPS1a'}$ point for process $e^+e^-\to h^0H^+
W^-$ as the functions of c.m.s energy in the energy range from
$900~{\rm GeV}$ to $1.8~{\rm TeV}$ are illustrated in Fig.3(a).
The tree-level total cross section is drawn in solid line, while
the full one-loop EW corrected cross section in dashed line. We
can read from the figure that at the position of
$\sqrt{s}=1.3~{\rm TeV}$, the Born and EW corrected cross sections
are $0.019~{\rm fb}$ and $0.015~{\rm fb}$, respectively. The
corresponding relative corrections are depicted in Fig.3(b) and we
can see that the absolute relative correction is very large when
the colliding energy goes close to the vicinity of the threshold
for process $e^+e^-\to h^0H^+ W^-$. Fig.3(b) shows that when
$\sqrt{s}=900~{\rm GeV}$, $\delta_{total}$ is about $-36\%$, but
when the $\sqrt{s}$ goes to $1.3~TeV$, the absolute relative
correction declines to the value of $19.2\%$.

\begin{figure}[htb]
\centering
\includegraphics[scale=0.4]{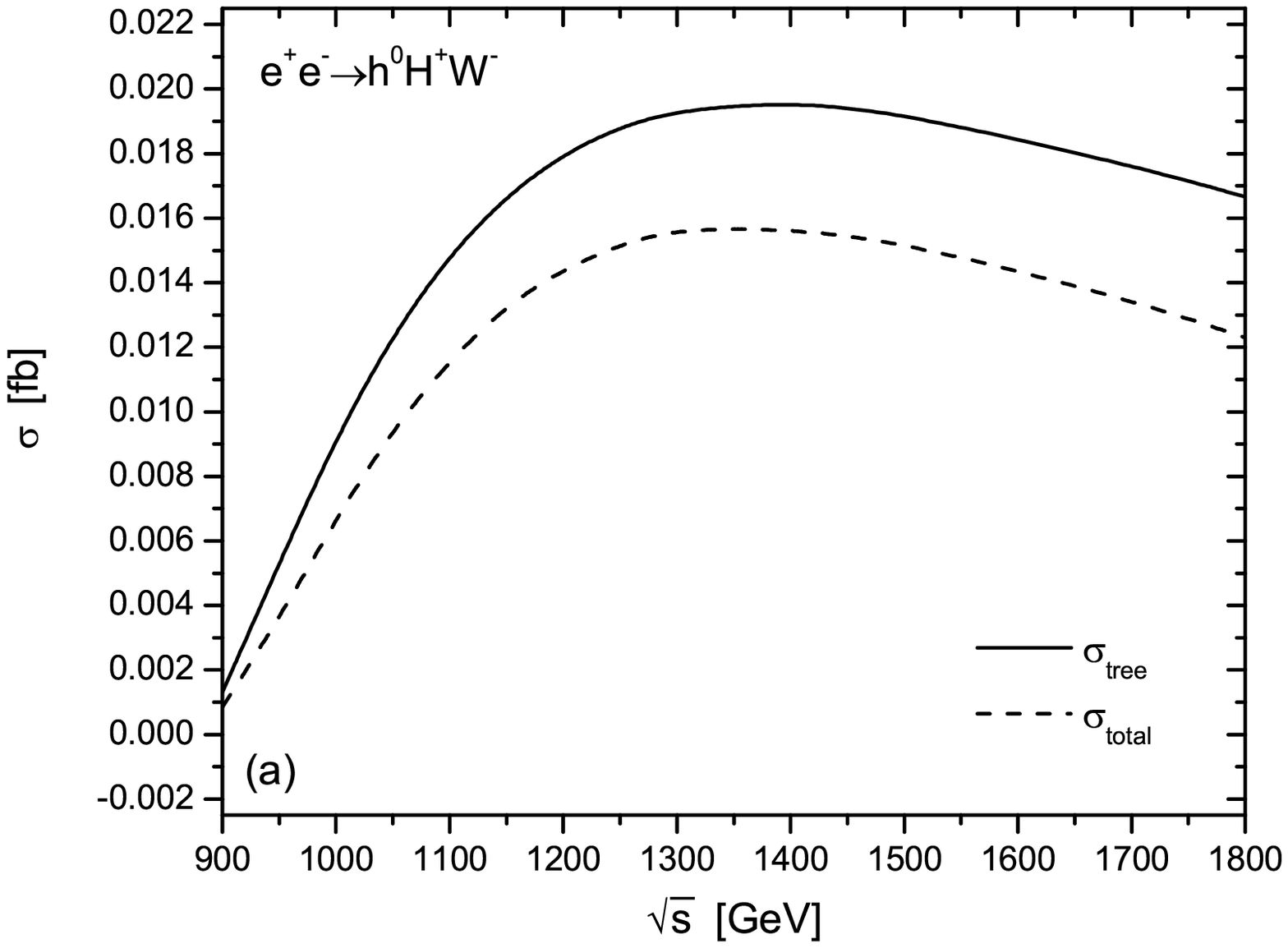}
\includegraphics[scale=0.4]{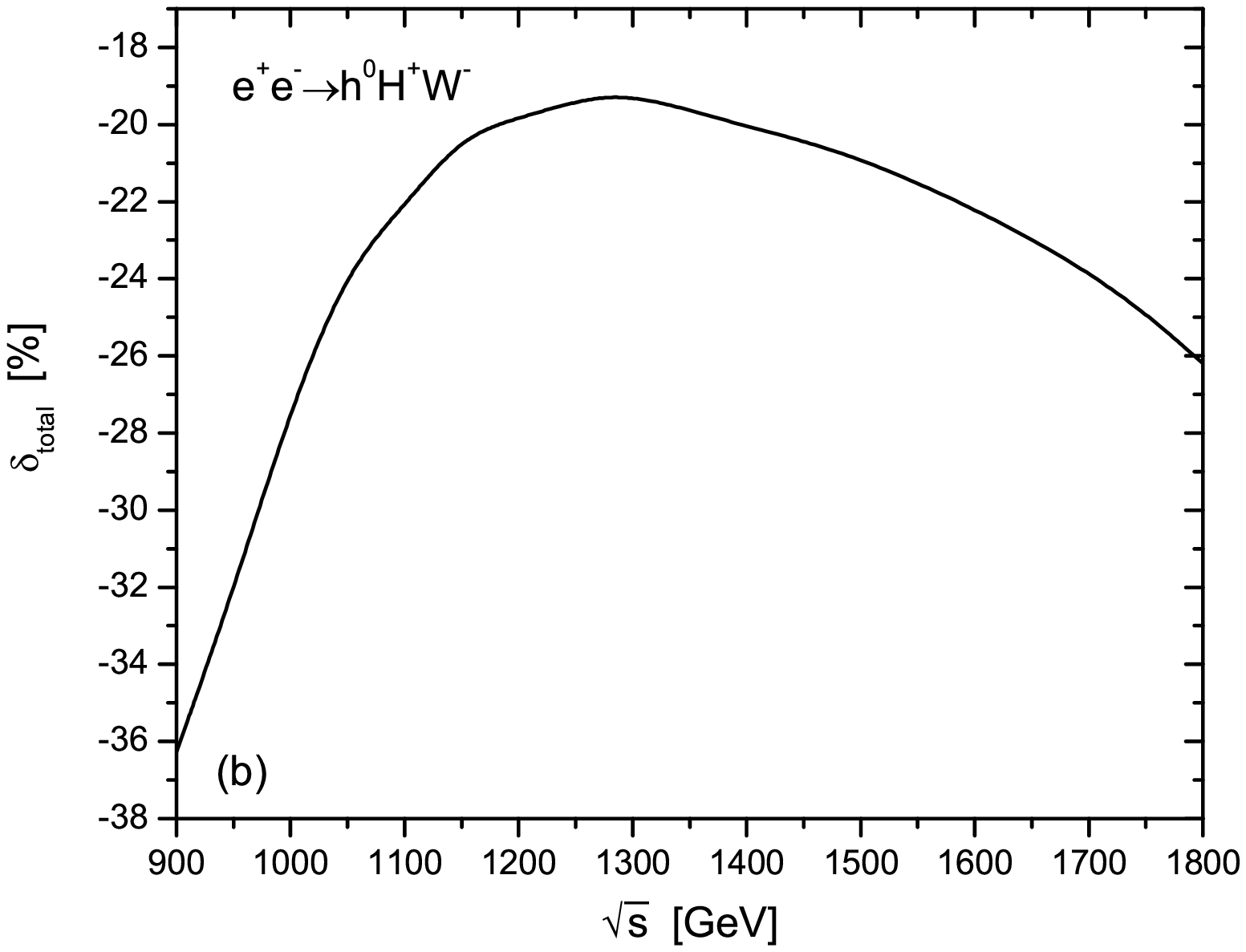}
\caption{The Born and full ${\cal O}(\alpha_{ew})$ electroweak
corrected cross sections as well as the corresponding relative
corrections $\delta_{total}$ as the functions of the c.m.s energy
$\sqrt{s}$ at ${\rm SPS1a'}$ point for process $e^+e^-\to h^0H^+
W^-$. (a) cross section versus $\sqrt{s}$. (b) $\delta_{total}$
versus $\sqrt{s}$.}
\end{figure}

\par
In Fig.4(a), we present the Born cross section $\sigma_{tree}$ and
one-loop electroweak corrected cross section $\sigma_{total}$ with
$M_{A^0}=200~{\rm GeV}$ and related MSSM parameters being taken
from ${\rm SPS1a'}$ scenario except Higgs sector for reaction
$e^+e^-\to h^0H^+ W^-$. In this case the mass of the lightest
neutral Higgs boson $M_{h^0}$ is equal to $117.2~GeV$, which is
fixed in the MSSM by using FormCalc program with the input
parameters: $M_{A^0}=200~GeV$ and OS $\tan \beta=10.31$. In
contrast with the plots in Fig.3(a-b), the complete EW one-loop
corrections enhance Born cross sections as shown in Fig.4(a-b),
and both the $\sigma_{tree}$ and the EW corrected $\sigma_{total}$
in Fig.4(a) are at least one order larger than in Fig.3(a).
$\sigma_{total}$ can reach maximal value $0.33~{\rm fb}$ at c.m.s
energy near $700~{\rm GeV}$ and decreases smoothly with the
increment of $\sqrt{s}$. The corresponding relative correction is
depicted in Fig.4(b) in the same parameter space, the correction
varies gently from $9\%$ to $21.4\%$.

\begin{figure}[htb]
\centering
\includegraphics[scale=0.4]{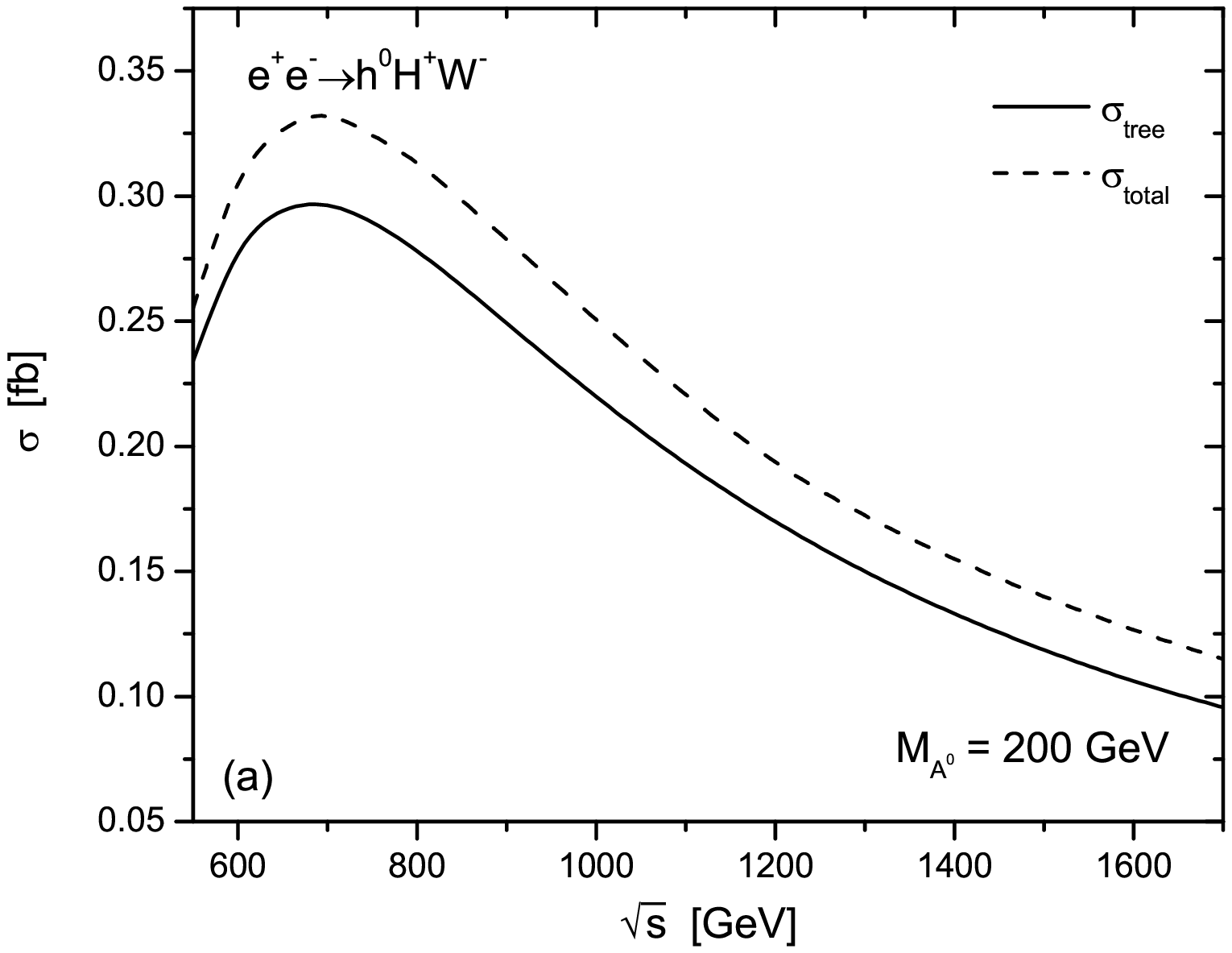}
\includegraphics[scale=0.4]{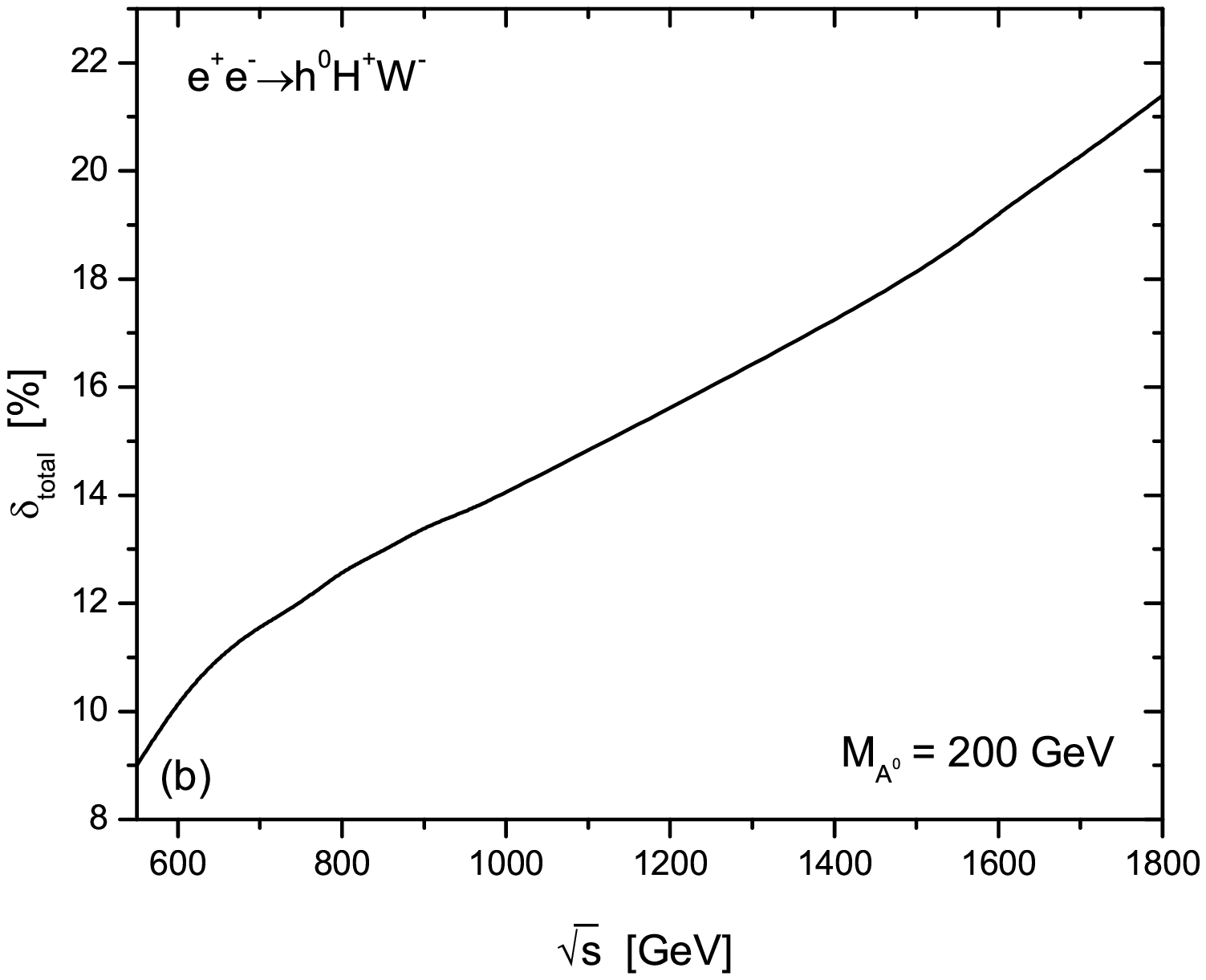}
\caption{The Born and full ${\cal O}(\alpha_{ew})$ electroweak
corrected cross sections as well as the corresponding relative
corrections $\delta_{total}$ as the functions of the c.m.s energy
$\sqrt{s}$ with $M_{A^0}=200~{\rm GeV}$ and related MSSM
parameters being taken from the ${\rm SPS1a'}$ scenario except
Higgs sector for process $e^+e^-\to h^0H^+ W^-$. (a) cross section
versus $\sqrt{s}$, (b) $\delta_{total}$ versus $\sqrt{s}$.}
\end{figure}

\par
In Fig.5(a-b) we depict the Born and EW corrected cross sections
and corresponding relative corrections as the functions of mass
$M_{A^0}$ with $\sqrt{s}=1000~{\rm GeV}$ and the relevant MSSM
parameters adopting from the ${\rm SPS1a'}$ point except Higgs
sector for process $e^+e^-\to h^0H^+ W^-$. The other Higgs boson
masses are determined in the MSSM by using FormCalc program with
$M_{A^0}$ and OS $\tan \beta=10.31$ being its input parameters.
The values of $M_{h^0}$ and $M_{H^+}$ corresponding to different
$M_{A^0}$ values are also scaled on x-axes in Fig.5(a-b). In
Fig.5(a), we see there is an abrupt change on each curves due to
the resonance effect of $H^+\to h^0 W^+$ at the position around
$M_{A^0}\sim 180 {\rm GeV}$(where $M_{h^0}=116.8~{\rm GeV}$ and
$M_{H^+}=197.2~{\rm GeV}$). After reaching its maximal value at
position of $M_{A^0}=210~{\rm GeV}$, the Born electroweak
corrected cross sections decrease gently. The dependence of
relative correction on $M_{A^0}$(or $M_{h^0}$, $M_{H^+}$) is
displayed in Fig.5(b). $\delta_{total}$ curve bends downward in
the vicinity of $M_{A^0} \sim 180~{\rm GeV}$ because of resonance
effect, and then decreases as $M_{A^0}$ goes up. The relative
correction varies from positive to negative value region and the
absolute value exceeds $10\%$ for both light and heavy Higgs
bosons.

\begin{figure}[htb]
\centering
\includegraphics[scale=0.4]{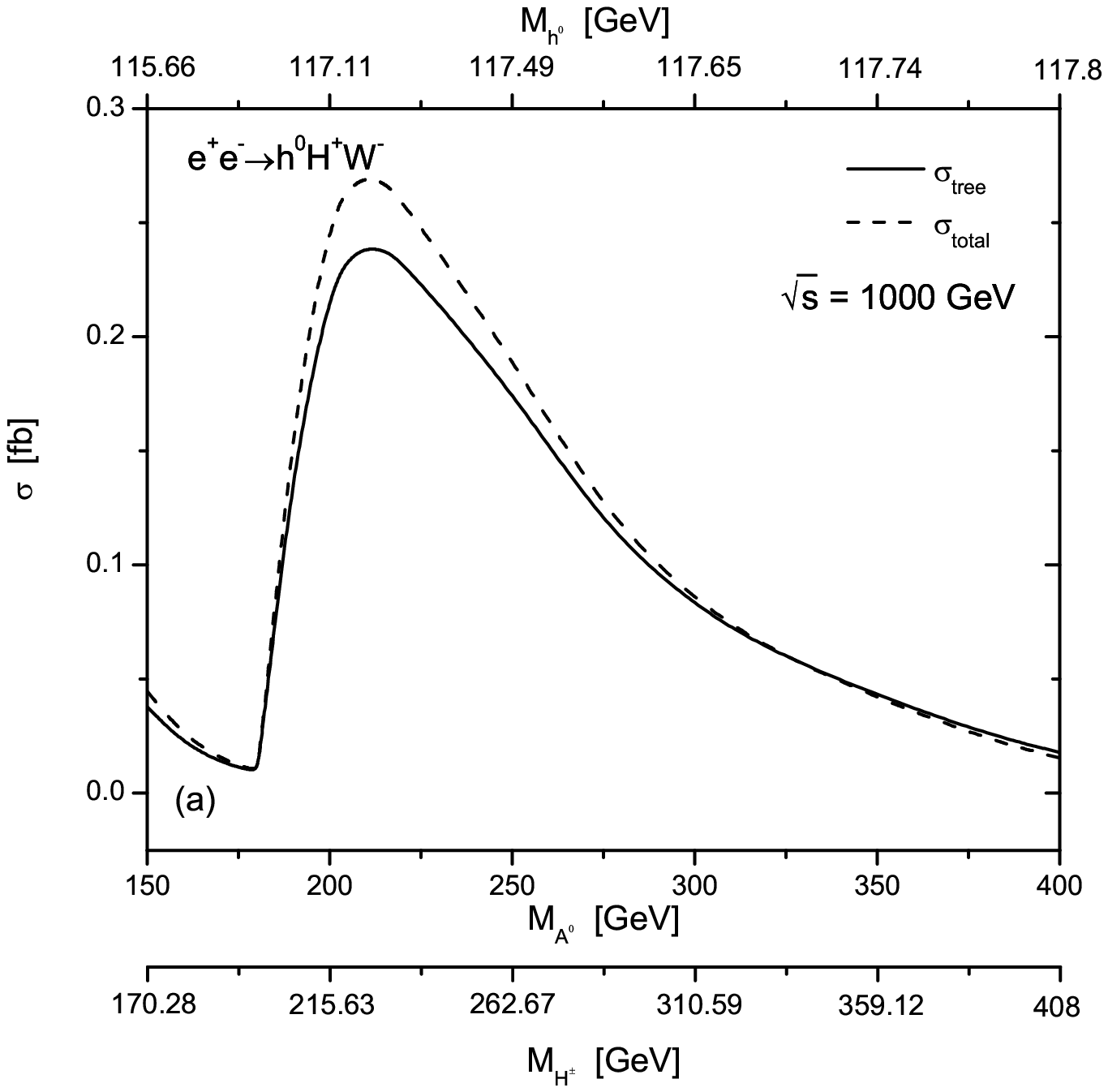}
\includegraphics[scale=0.4]{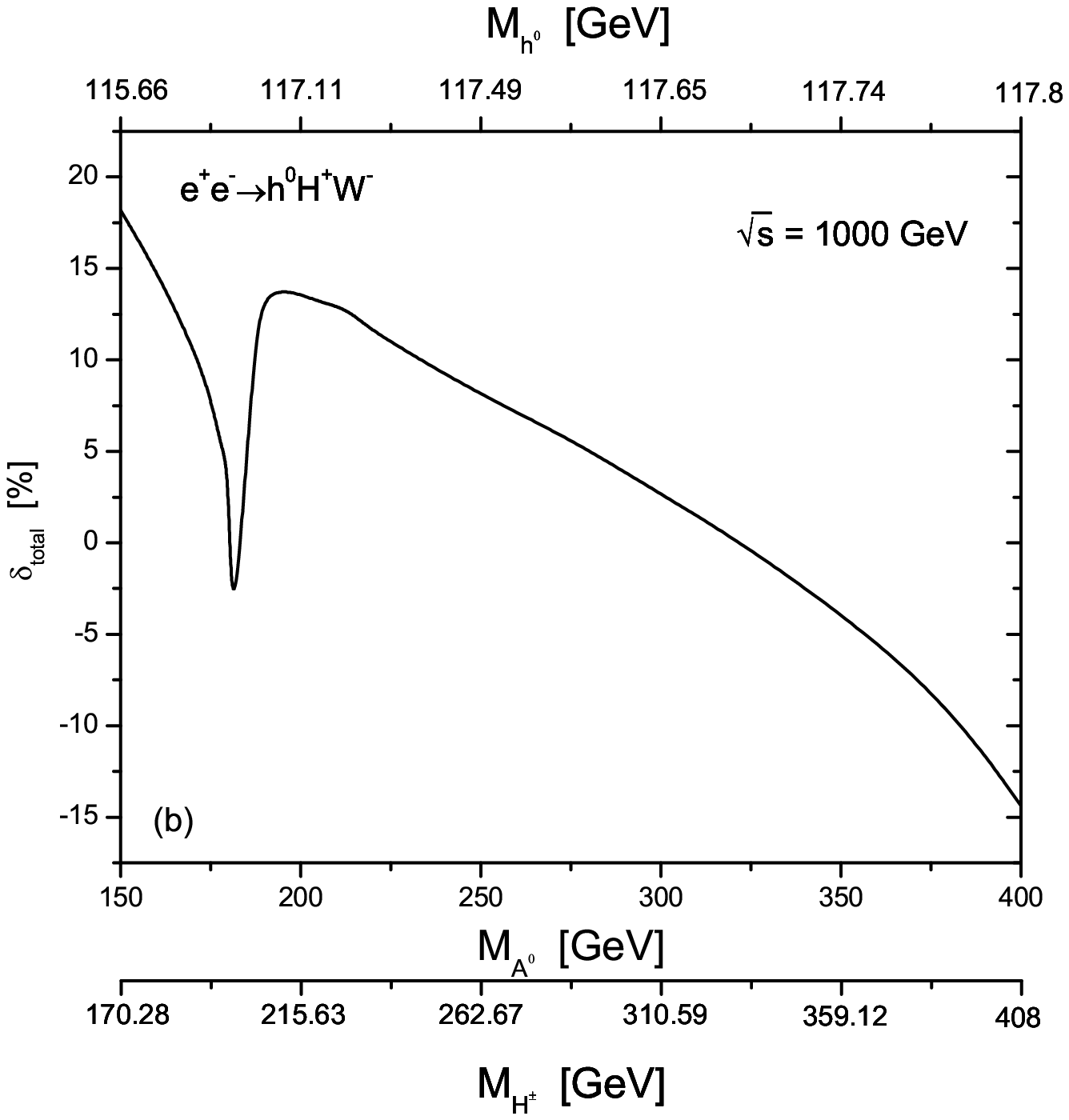}
\caption{The Born and full ${\cal O}(\alpha_{ew})$ electroweak
corrected cross sections as well as the corresponding EW relative
corrections $\delta_{total}$ as the functions of $M_{A^0}$ with
$\sqrt{s}=1000~{\rm GeV}$ and relevant MSSM parameters taken from
${\rm SPS1a'}$ point except Higgs sector being obtained by running
FormCalc program with input parameters: $M_{H^0}$ and OS
$\tan\beta=10.31$, for process $e^+e^-\to h^0H^+ W^-$. (a) cross
section versus $M_{A^0}$, (b) relative correction versus
$M_{A^0}$. }
\end{figure}

\par
We depict the Born and full one-loop electroweak corrected cross
sections as the functions of c.m.s energy $\sqrt{s}$ with the MSSM
parameters at the ${\rm SPS1a'}$ point in Fig.6(a) and Fig.6(b)
for process $e^+e^-\to H^0H^+ W^-$ and $e^+e^-\to A^0H^+W^-$,
separately. In the ${\rm SPS1a'}$ scenario, the value of heavier
neutral and charged Higgs boson masses($M_{H^0}$, $M_{A^0}$ and
$M_{H^+}$) have the values beyond $420~{\rm GeV}$, accordingly the
threshold energy of this parameter choice can reach ${\rm 950~
GeV}$. Therefore, in Fig6(a-b) we plot the c.m.s. energy
$\sqrt{s}$ in the range between $1000~{\rm GeV}$ and $1800~{\rm
GeV}$. We see both of Born and ${\cal O}(\alpha_{ew})$ corrected
cross sections increase with the increment of the colliding
energy. Fig.6(c) and Fig.6(d) show the dependence of the relative
corrections $\delta_{total}$ of processes $e^+e^-\to H^0H^+W^-$
and $e^+e^-\to A^0H^+W^-$ on the c.m.s energy $\sqrt{s}$
respectively, with the same MSSM parameters as taken in
Fig.6(a-b). There we see that when the colliding energy $\sqrt{s}$
goes up, both curves in Fig.6(c-d) for relative correction
$\delta_{total}$ increase rapidly in the vicinity of threshold
energy, and then becomes increasing smoothly with $\delta_{total}$
value being in the range of $[-16\%, -23\%]$ when c.m.s energy is
beyond $1.2~{\rm TeV}$.

\begin{figure}[htp]
\centering
\includegraphics[scale=0.4]{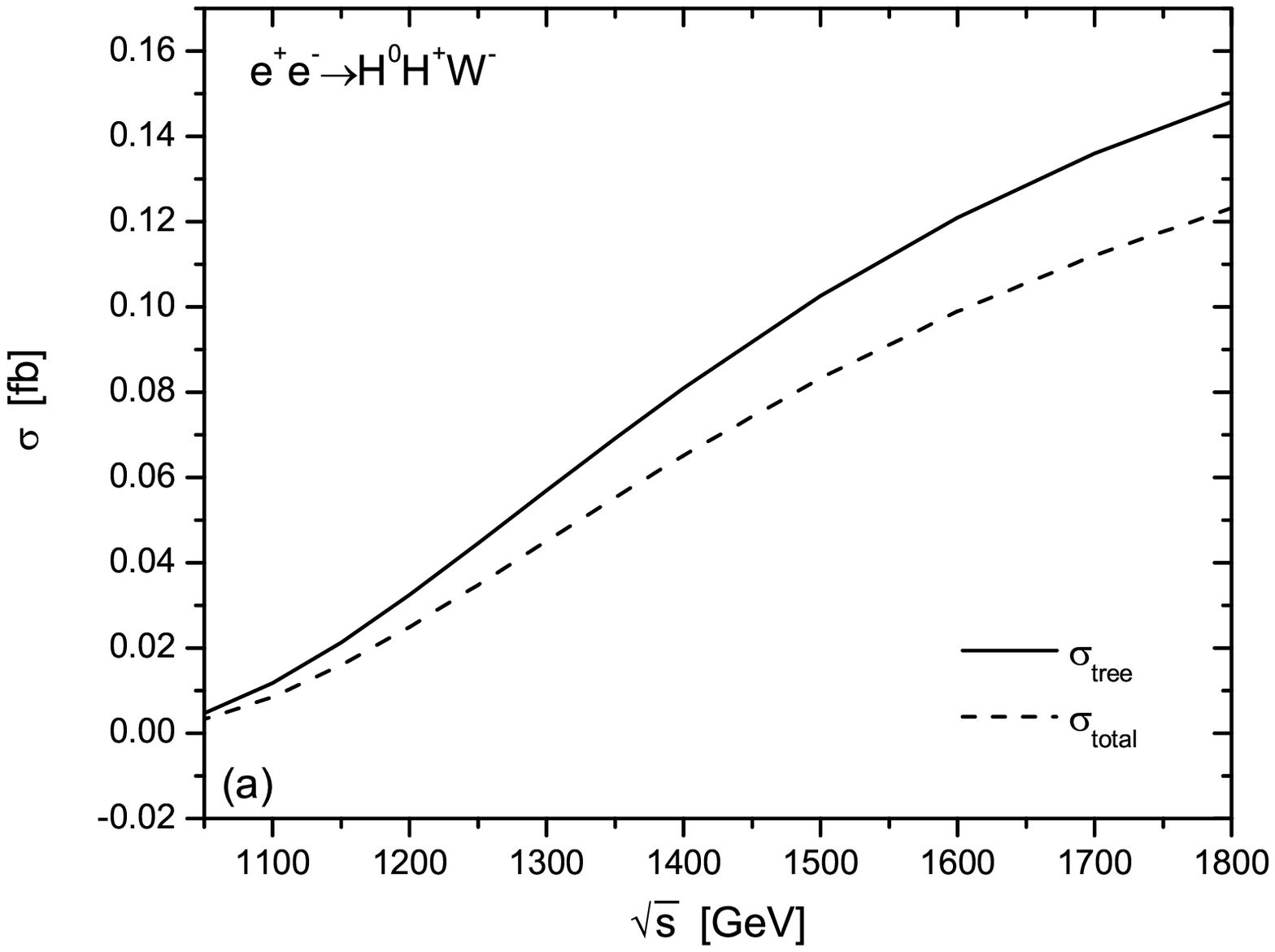}
\includegraphics[scale=0.4]{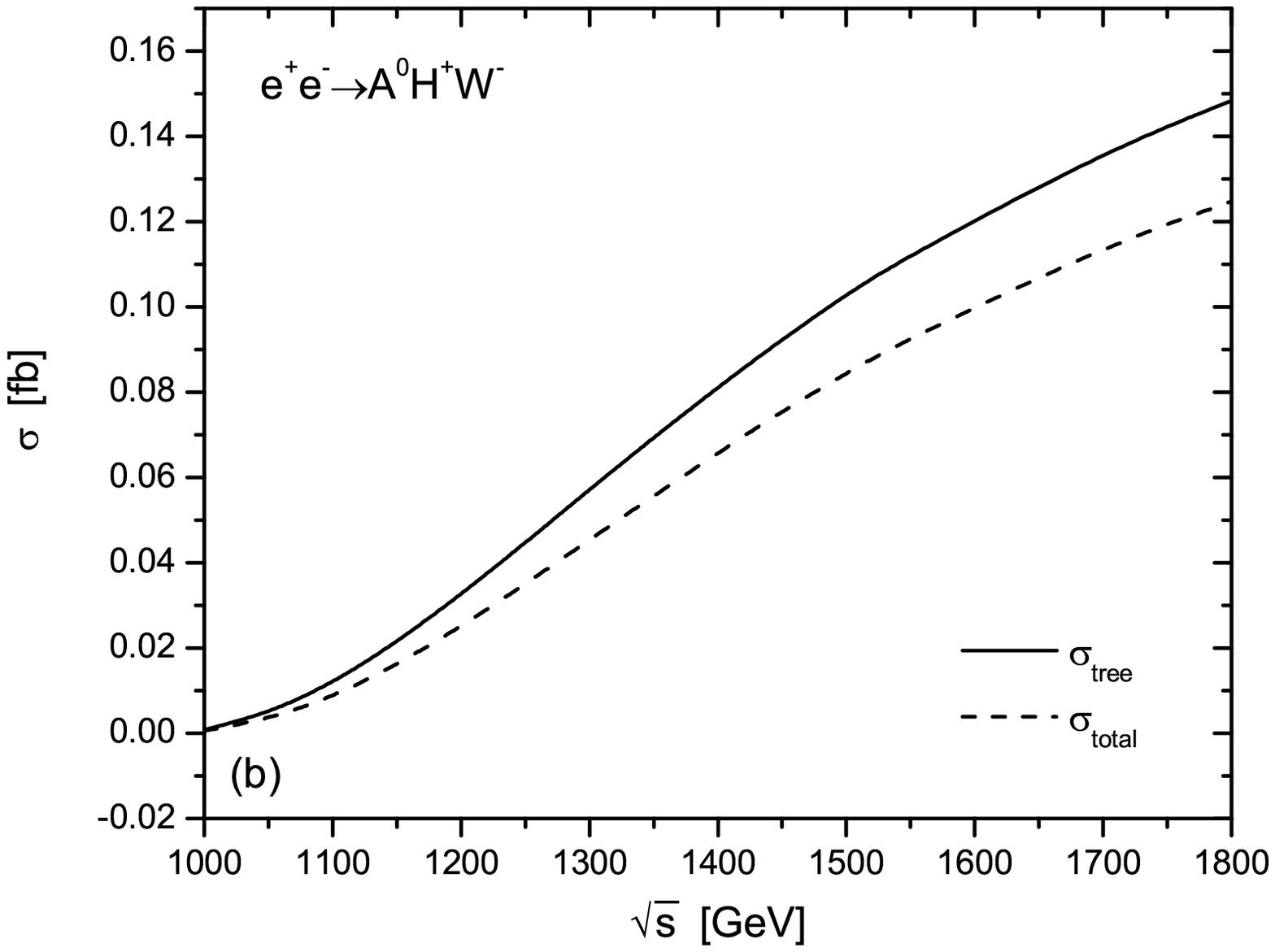}
\includegraphics[scale=0.4]{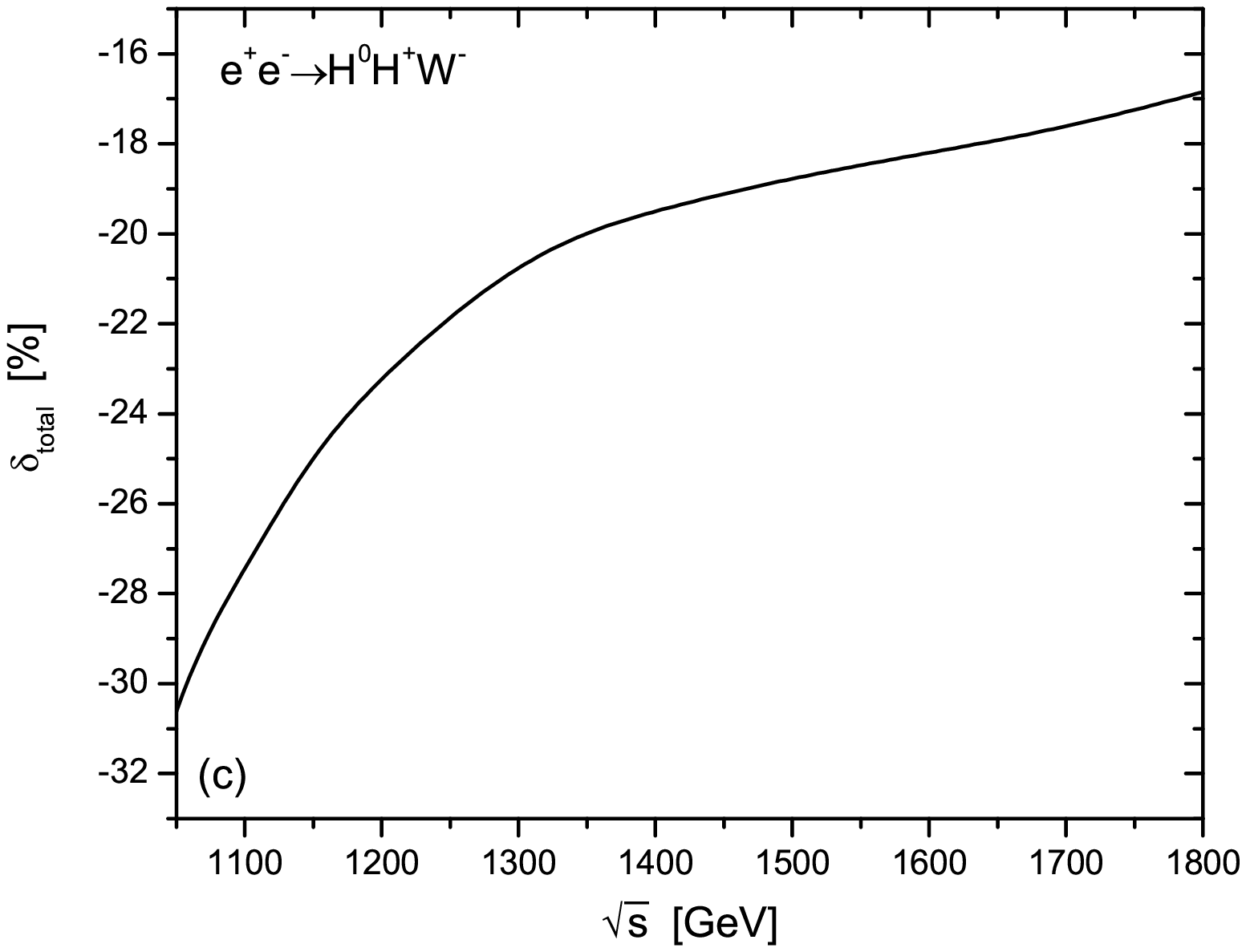}
\includegraphics[scale=0.4]{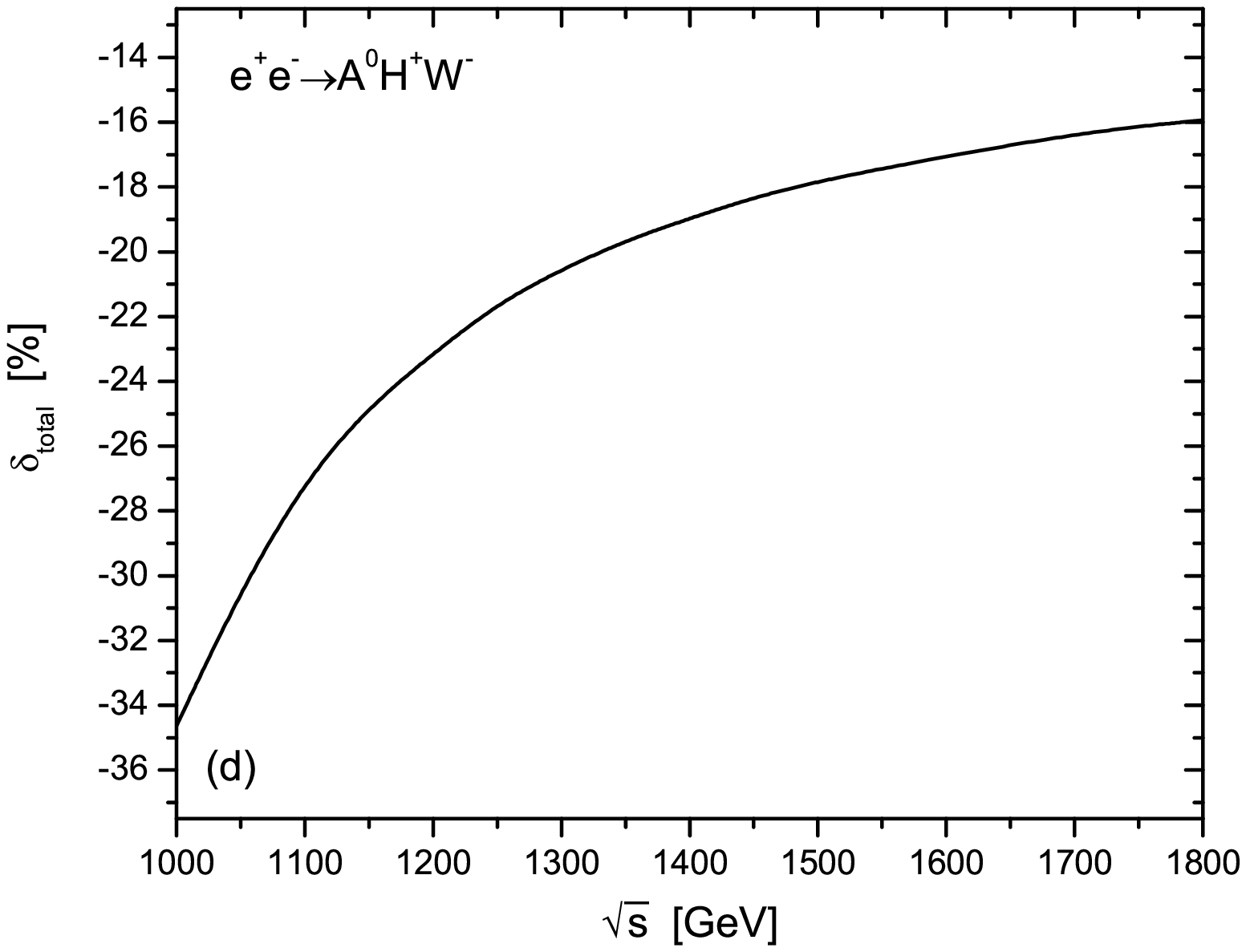}
\caption{The Born and full ${\cal O}(\alpha_{ew})$ electroweak
corrected cross sections as well as the corresponding relative
corrections $\delta_{total}$ as the functions of the c.m.s energy
$\sqrt{s}$ in the ${\rm SPS1a'}$ scenario for process $e^+e^-\to
H^0(A^0) H^+ W^-$. (a) cross section versus $\sqrt{s}$ for
$e^+e^-\to H^0 H^+ W^-$, (b) cross section versus $\sqrt{s}$ for
$e^+e^-\to A^0 H^+ W^-$, (c) $\delta_{total}$ versus $\sqrt{s}$
for $e^+e^-\to H^0 H^+ W^-$, (d)$\delta_{total}$ versus $\sqrt{s}$
for $e^+e^-\to A^0 H^+ W^-$. }
\end{figure}

\par
The Born and full ${\cal O}(\alpha_{ew})$ corrected cross sections
as the functions of c.m.s energy $\sqrt{s}$ with $M_{A^0}=200~{\rm
GeV}$ and relevant ${\rm SPS1a'}$ MSSM parameters except Higgs
sector for processes $e^+e^-\to H^0 H^+ W^-$ and $e^+e^-\to A^0
H^+ W^-$ are plotted in Fig.7(a) and Fig.7(b), separately. Again,
the values of other Higgs boson masses are obtained from running
FormCalc program with input parameter values of $M_{A^0}=200~{\rm
GeV}$ and OS $\tan \beta=10.31$. Both Born cross section and
corrected section increase gently when $\sqrt{s}$ goes beyond the
threshold, while when $\sqrt{s}$ approaches to $1.2~{\rm TeV}$,
the ${\cal O}(\alpha_{ew})$ EW corrected cross section can achieve
the maximal value of about $0.4~{\rm fb}$. The corresponding EW
relative corrections with the same conditions as in Fig.7(a) and
Fig.7(b), are described in Fig.7(c) and (d), respectively. As we
can see in these figures, the line-shapes of Born and EW corrected
cross sections(relative correction) for $e^+e^-\to H^0 H^+ W^-$
process are very similar with the corresponding ones for
$e^+e^-\to A^0 H^+ W^-$ process.

\begin{figure}[htp]
\centering
\includegraphics[scale=0.4]{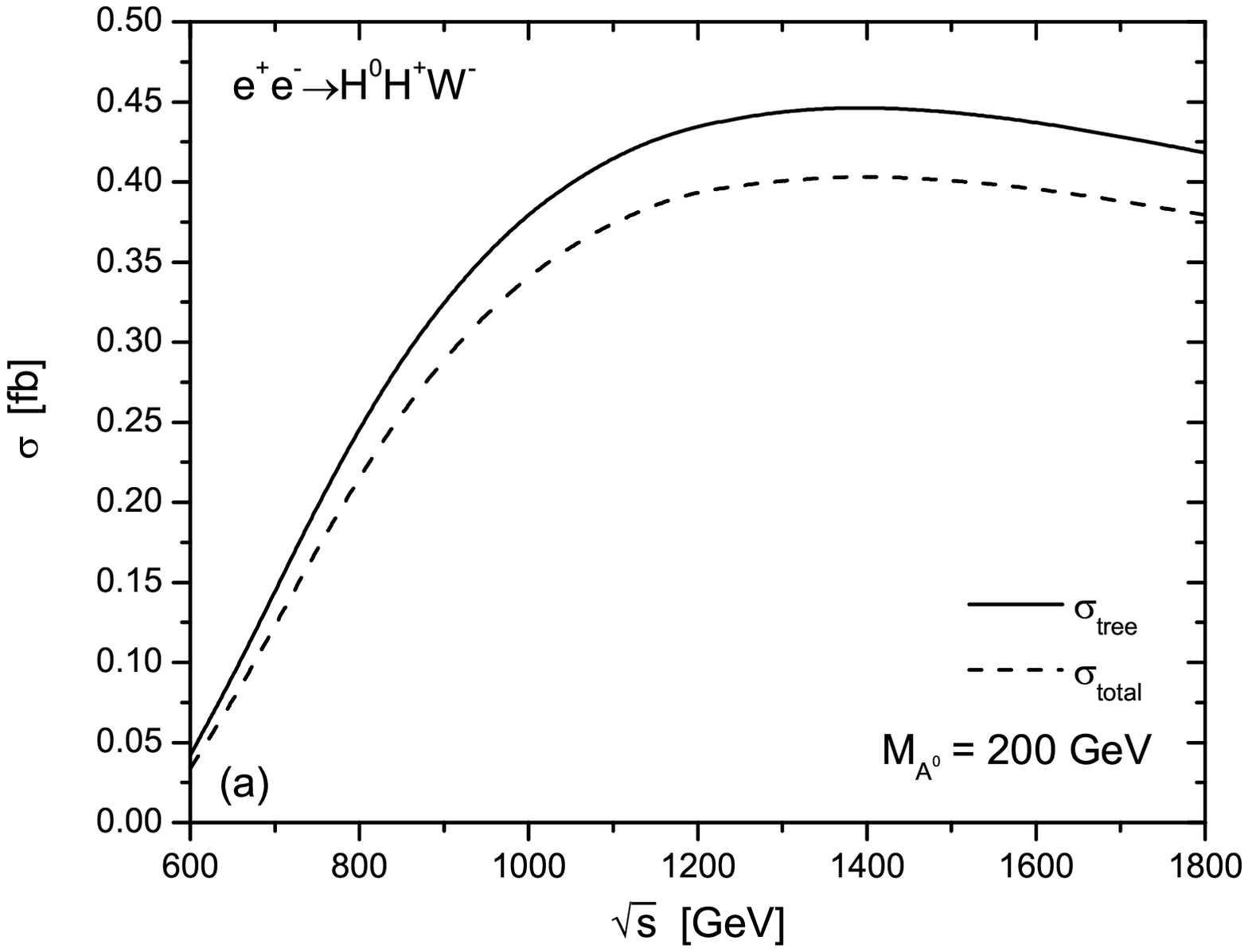}
\includegraphics[scale=0.4]{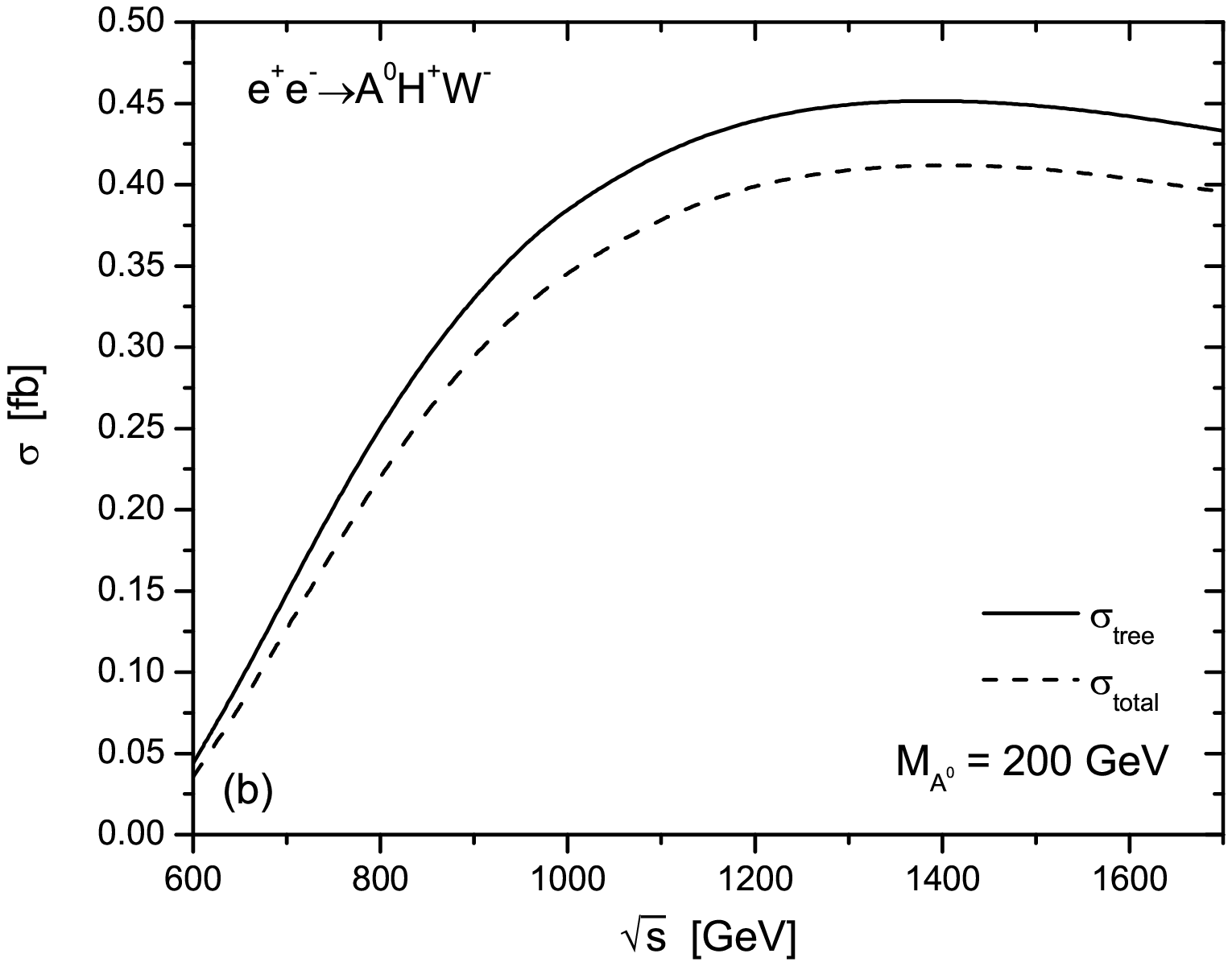}
\includegraphics[scale=0.4]{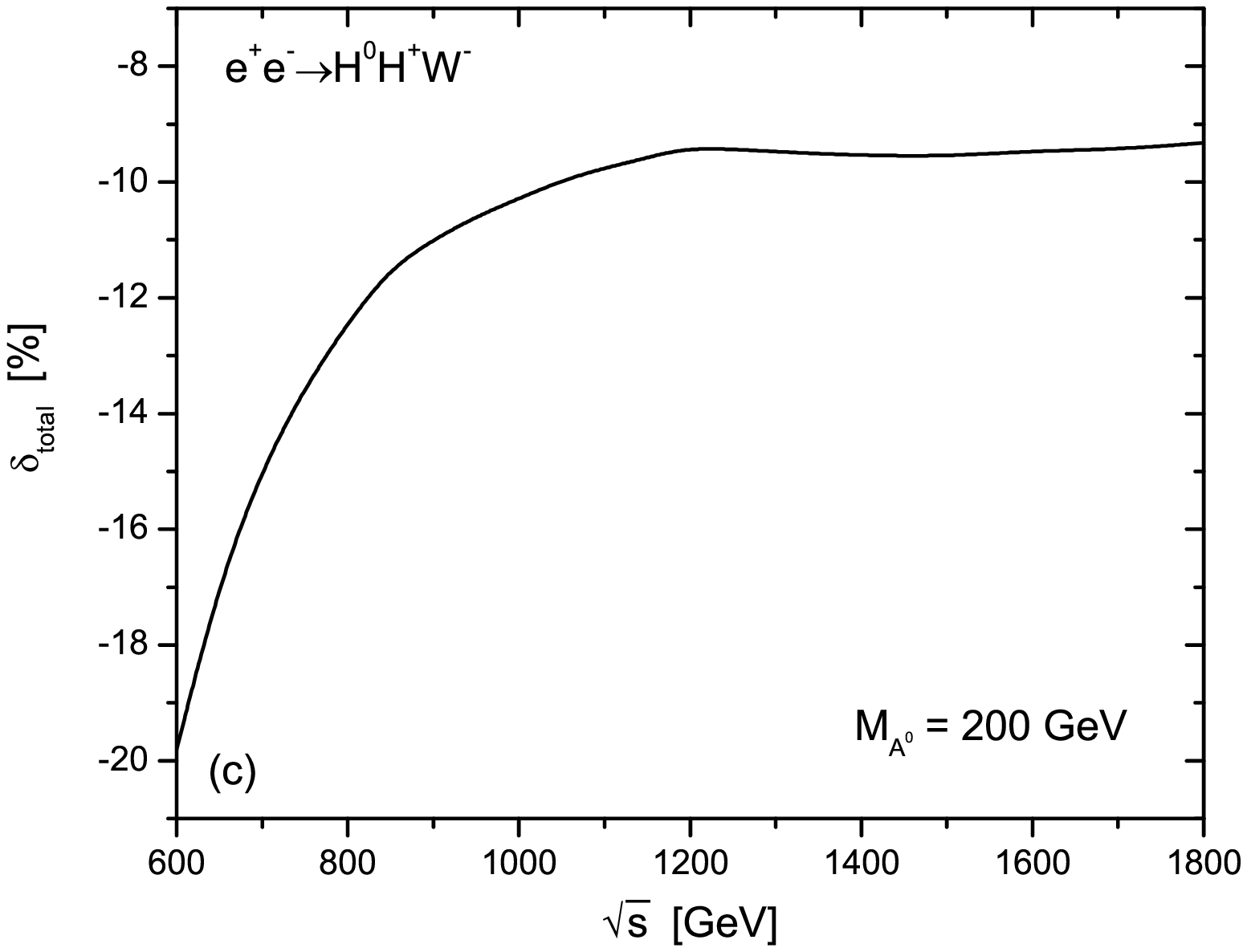}
\includegraphics[scale=0.4]{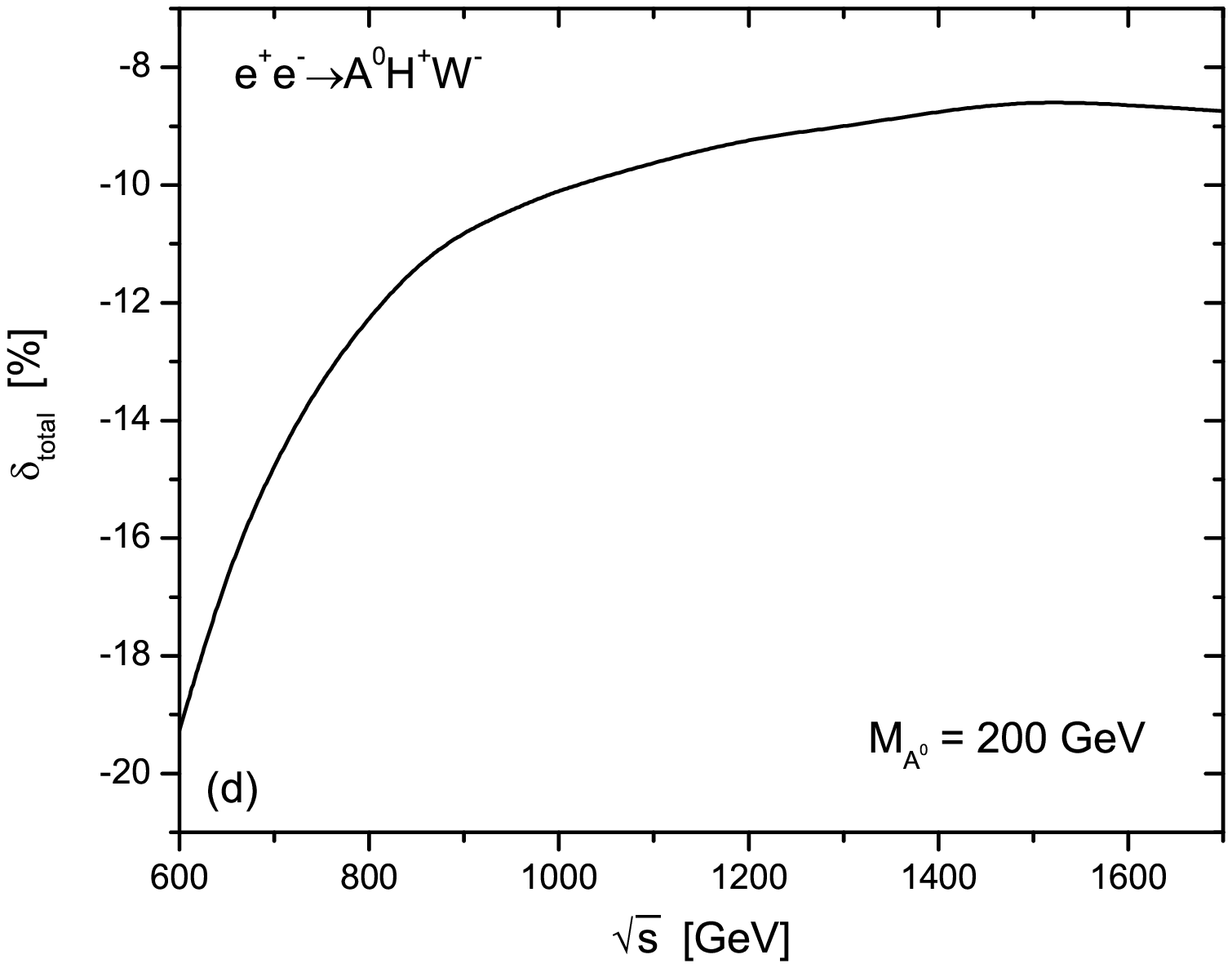}
\caption{The Born and full ${\cal O}(\alpha_{ew})$ electroweak
corrected cross sections as well as the corresponding relative
corrections $\delta_{total}$ as the functions of the c.m.s energy
$\sqrt{s}$ with $M_{A^0}=200~{\rm GeV}$ and relevant ${\rm
SPS1a'}$ MSSM parameters except Higgs sector for processes
$e^+e^-\to H^0(A^0) H^+ W^-$.(a) cross section versus $\sqrt{s}$
for $e^+e^-\to H^0 H^+ W^-$, (b) cross section versus $\sqrt{s}$
for $e^+e^-\to A^0 H^+ W^-$, (c) $\delta_{total}$ versus
$\sqrt{s}$ for $e^+e^-\to H^0 H^+ W^-$, (d)$\delta_{total}$ versus
$\sqrt{s}$ for $e^+e^-\to A^0 H^+ W^-$. }
\end{figure}

\par
The Born and EW corrected cross sections as the functions of input
parameter $M_{A^0}$ are drawn in Fig.8(a) for process $e^+e^-\to
H^0H^+ W^-$, and the corresponding relative corrections versus
$M_{A^0}$ are depicted in Fig.8(b), with c.m.s energy
$\sqrt{s}=1200~{\rm GeV}$ and the relative MSSM parameters taken
from ${\rm SPS1a'}$ point except the Higgs sector being fixed by
using FormCalc program. In Fig.8(a) the one-loop electroweak
corrected cross section decreases from $0.56~{\rm fb}$ to
$0.05~{\rm fb}$ with the increment of $M_{H^+}(M_{H^0})$ from
$170.28~GeV(152.96~GeV)$ to $408~GeV(400.51~GeV)$ for process
$e^+e^-\to H^0H^+ W^-$. Fig.8(b) shows the relative correction for
process $e^+e^-\to H^0H^+ W^-$ goes from $-7.8\%$ to $-20.6\%$
when $M_{H^+}(M_{H^0})$ goes up from $170.28~GeV(152.96~GeV)$ to
$408~GeV(400.51~GeV)$. For channel $e^+e^-\to A^0H^+ W^-$, the
cross section and EW relative correction versus $M_{A^0}$ are
depicted in Fig.8(c,d) respectively, with $\sqrt{s}$ being $\rm
{1000~GeV}$ and other MSSM parameters taken from ${\rm SPS1a'}$
point except Higgs sector being determined by using FormCalc
package. We can read from Fig.8(c-d) that the EW corrected cross
section goes down from $0.56~{\rm fb}$ to $0.005~{\rm fb}$ and the
relative correction varies from $-7\%$ to $-28.6\%$, when input
value of $M_{A^0}$ increases from $150~GeV$ to $400~GeV$. In these
four figures of Fig.8, we can see the behavior of these two
processes $e^+e^-\to H^0(A^0)H^+ W^-$ are very similar except tiny
different in the numerical value.

\begin{figure}[htp]
\centering
\includegraphics[scale=0.45]{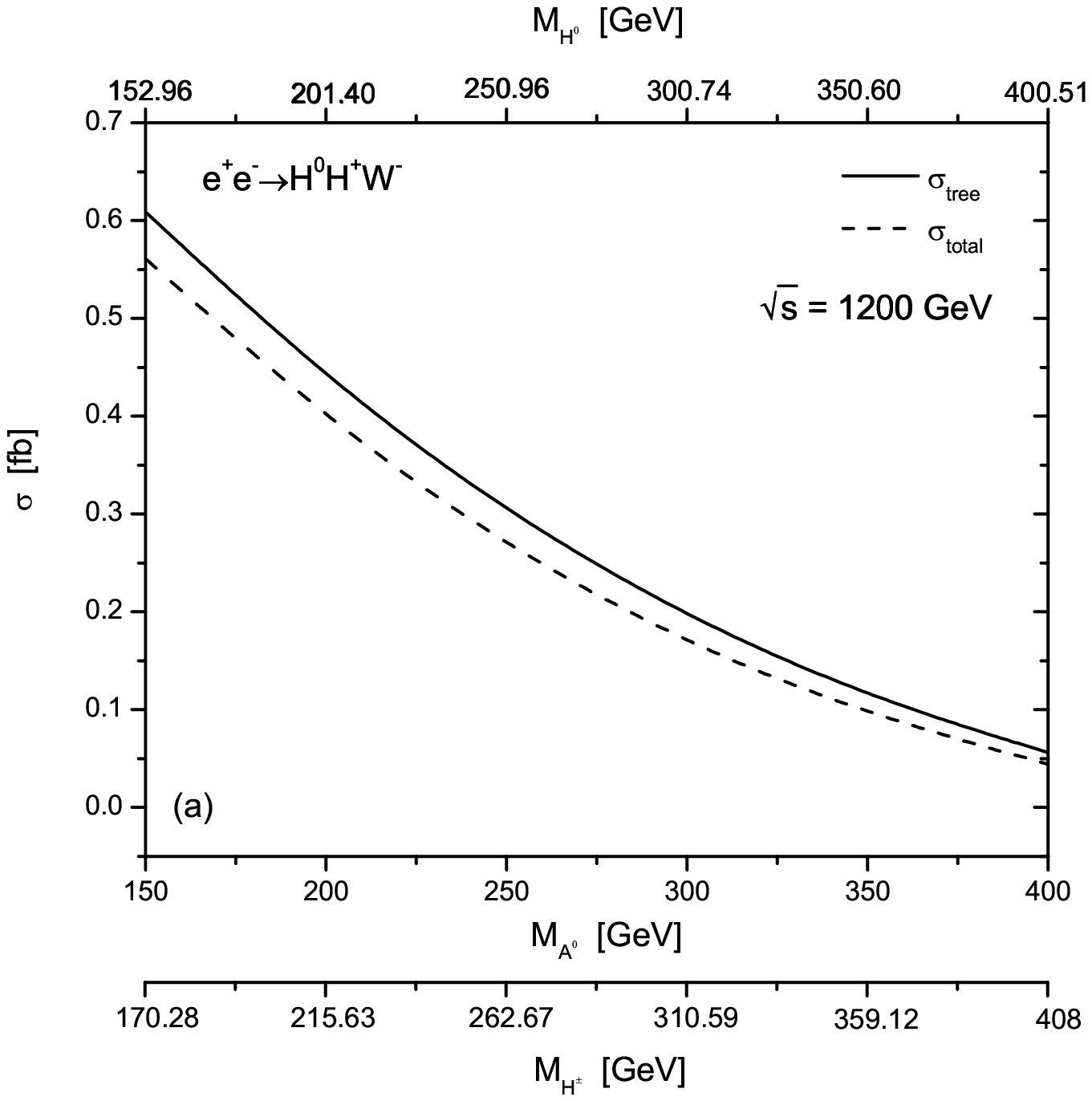}
\includegraphics[scale=0.45]{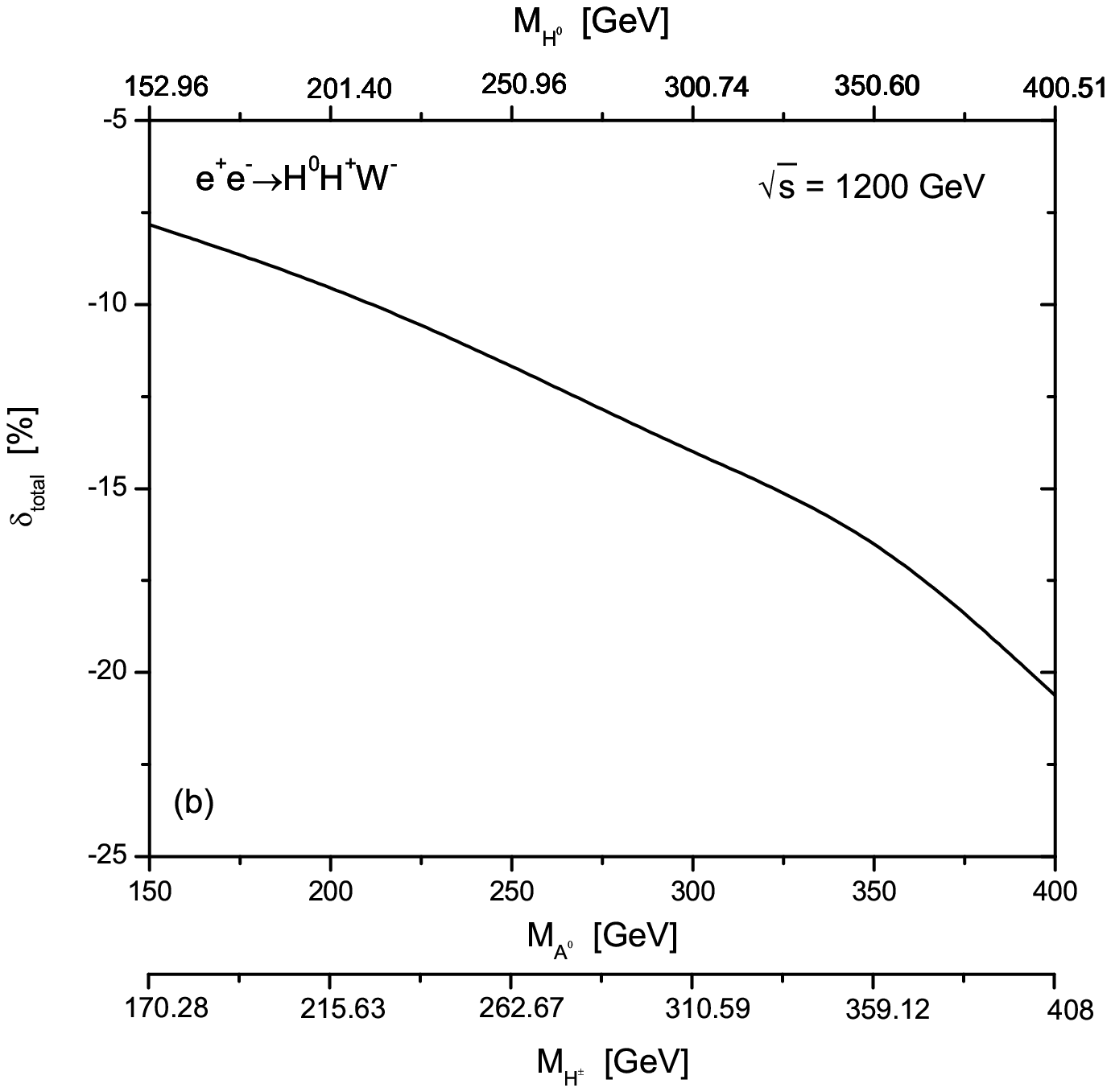}
\includegraphics[scale=0.45]{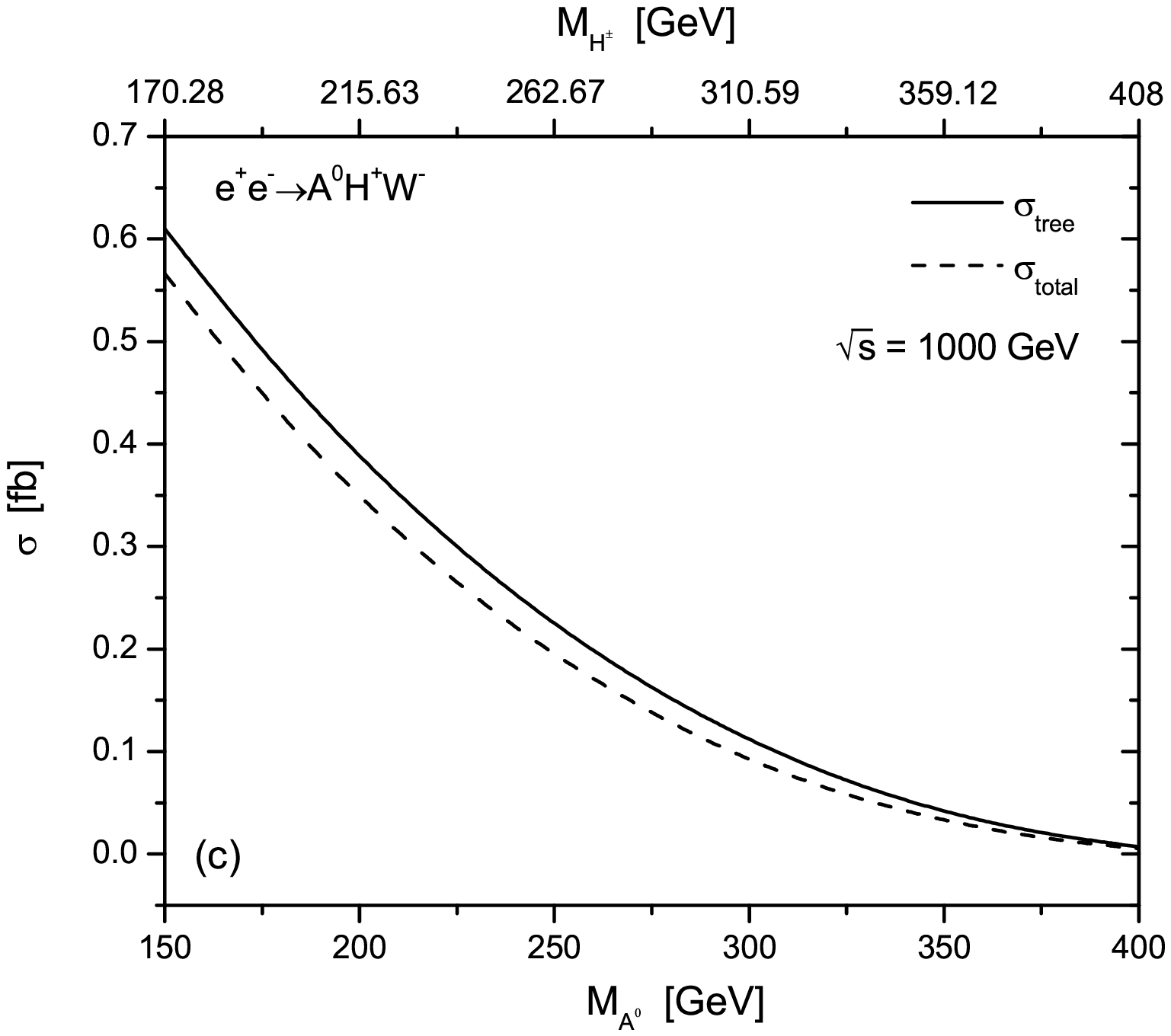}
\includegraphics[scale=0.45]{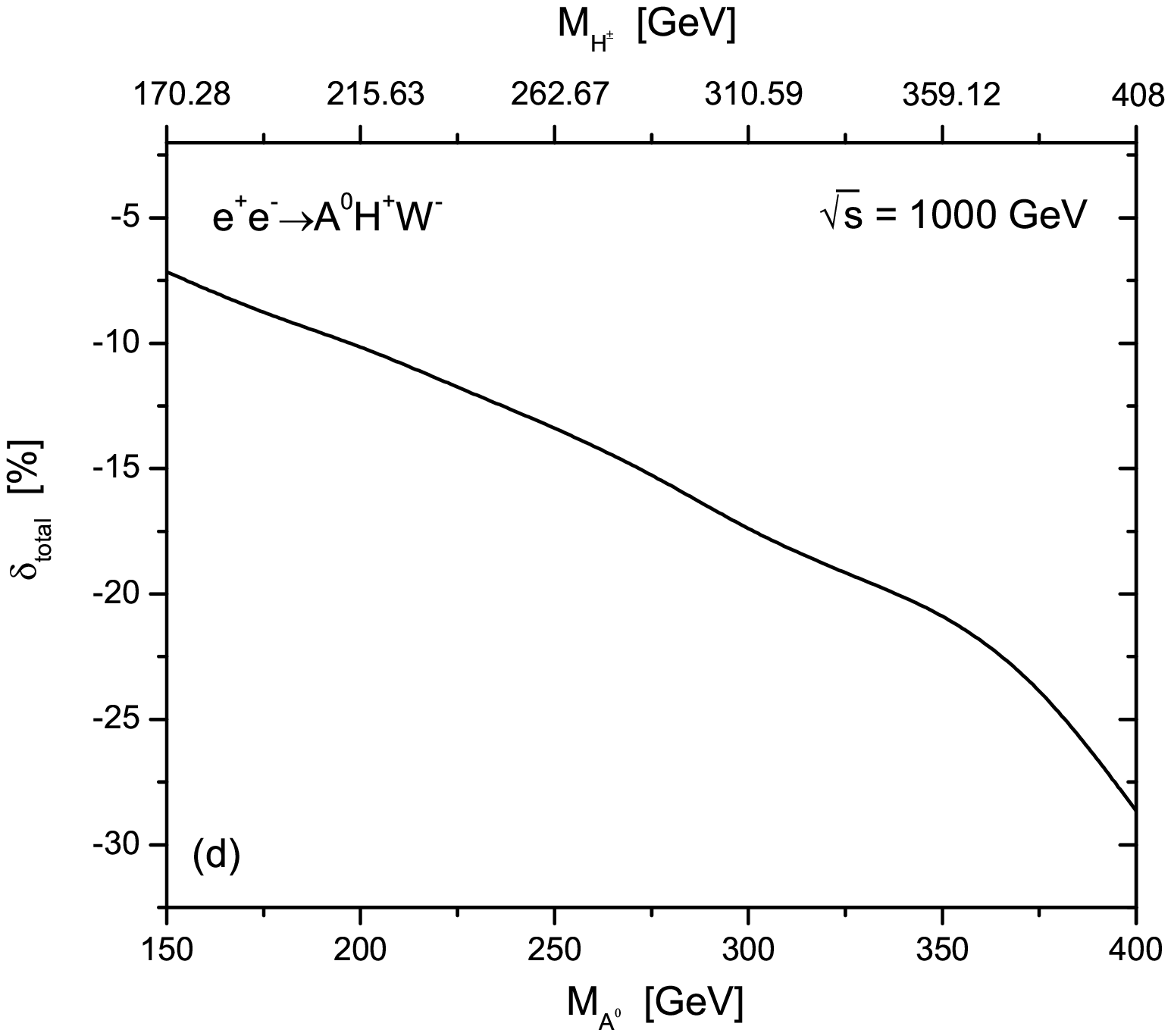}
\caption{The Born and full ${\cal O}(\alpha_{ew})$ electroweak
corrected cross sections as well as the corresponding relative
corrections $\delta_{total}$ as the functions of $M_{A^0}$ with
c.m.s energy $\sqrt{s}=1200~{\rm GeV}$ and relevant MSSM
parameters in ${\rm SPS1a'}$ except Higgs sector, for processes
$e^+e^-\to H^0(A^0) H^+ W^-$. (a) cross section versus $M_{A^0}$
for $e^+e^-\to H^0 H^+ W^-$, (b) $\delta_{total}$ versus $M_{A^0}$
for $e^+e^-\to H^0 H^+ W^-$, (c) cross section versus $M_{A^0}$
for $e^+e^-\to A^0 H^+ W^-$, (d)$\delta_{total}$ versus $M_{A^0}$
for $e^+e^-\to A^0 H^+ W^-$.}
\end{figure}

\par
From above numerical results we can see that the experimental
statistic errors for processes $e^+e^-\to h^0H^+W^-$, $e^+e^-\to
H^0H^+W^-$ and $e^+e^-\to A^0H^+W^-$ can reach $5.5\%$, $4.2\%$
and $4.2\%$, respectively. If we assume $\sqrt{s}=0.6-1.8~TeV$,
$\int Ldt \sim 1000~fb^{-1}$, and do not consider the backgrounds
to the $e^+e^-\to \Phi^0H^+W^-(\Phi=h^0,H^0,A^0)$ signals and the
systematic errors of the experiment(due to lacking their studies
at ILC), the EW corrections to the cross sections of these three
processes should be taken into account at a linear collider with
realistic accuracy better than $10 \%$.

\section{Summary}
\par
In this paper, we present the calculation of the full ${\cal
O}(\alpha_{ew})$ electroweak corrections to the processes
$e^+e^-\to \Phi^0 H^\pm W^\mp(\Phi=h^0,H^0,A^0)$ at a LC in the
MSSM. In the numerical calculation, we generally adopt the MSSM
parameters at the reference point ${\rm SPS1a'}$ defined in the
SPA project. For channel $e^+e^-\to h^0H^\pm W^\mp$, we present
the details for treating the charged Higgs boson resonance in the
calculation up to one-loop level. Our numerical results show that
the EW relative correction for the process $e^+e^-\to h^0H^\pm
W^\mp$ can be either positive or negative with the increment of
Higgs boson mass $M_{A^0}$, and the relative correction varies in
the range of $[-15\%,~20\%]$ as the $M_{A^0}$ goes from $150~GeV$
to $400~GeV$. For channels $e^+e^-\to H^0(A^0)H^\pm W^\mp$, the EW
correction generally reduce the Born cross sections and the
relative correction is typically of order $-10\%\sim -20\%$. We
conclude that the complete ${\cal O}(\alpha_{ew})$ electroweak
corrections to the process $e^+e^-\to \Phi^0 H^\pm
W^\mp(\Phi=h^0,H^0,A^0)$ are generally significant and cannot be
neglected in the precise experiment analysis.

\vskip 5mm
\par
\noindent{\large\bf Acknowledgments:} This work was supported in
part by the National Natural Science Foundation of China, the
Education Ministry of China and a special fund sponsored by
Chinese Academy of Sciences.

\end{document}